\documentclass[fleqn,usenatbib]{mnras}

\usepackage[T1]{fontenc}

\DeclareRobustCommand{\VAN}[3]{#2}
\let\VANthebibliography\thebibliography
\def\thebibliography{\DeclareRobustCommand{\VAN}[3]{##3}\VANthebibliography}

\usepackage{graphicx}	
\usepackage{amsmath}	
\usepackage{amssymb}	

\usepackage{newtxtext,newtxmath}
\usepackage{scalefnt}
\usepackage{xspace}

\newcommand{\lya}{Ly$\alpha$}
\newcommand\smaller[2][0.85]{{\scalefont{#1}#2}}
\newcommand{\CRKHACC}{\smaller{CRK-HACC}\xspace}
\newcommand{\HACC}{\smaller{HACC}\xspace}

\title[Eulerian vs.~SPH for \lya\ statistics]{ Modeling the Lyman-$\alpha$ forest with Eulerian and SPH hydrodynamical methods}

\author[S. Chabanier et al.]{
Sol\`ene Chabanier$^{1}$\thanks{E-mail: schabanier@lbl.gov},
J.D. Emberson$^{2}$,
Zarija Luki\'c$^{1}$,
Jesus Pulido$^{3}$,
Salman Habib$^{2,4}$,
Esteban Rangel$^{2}$,
\newauthor
~Jean Sexton$^{1}$,
Nicholas Frontiere$^{2}$, and
Michael Buehlmann$^{4}$
\\
$^{1}$Lawrence Berkeley National Laboratory, Berkeley, CA 94720, USA\\
$^{2}$CPS Division, Argonne National Laboratory, Lemont, IL 60439, USA\\
$^{3}$Los Alamos National Laboratory, Los Alamos, NM 87545, USA\\
$^{4}$HEP Division, Argonne National Laboratory, Lemont, IL 60439, USA
}

\date{Accepted XXX. Received YYY; in original form ZZZ}

\pubyear{2022}

\begin{document}
\label{firstpage}
\pagerange{\pageref{firstpage}--\pageref{lastpage}}
\maketitle

\begin{abstract}
We compare two state-of-the-art numerical codes to study the overall accuracy in modeling the intergalactic medium and reproducing Lyman-$\alpha$ forest observables for DESI and high-resolution data sets. The codes employ different approaches to solving both gravity and modeling the gas hydrodynamics. The first code, Nyx, solves the Poisson equation using the Particle-Mesh (PM) method and the Euler equations using a finite volume method. The second code, \CRKHACC, uses a Tree-PM method to solve for gravity, and an improved Lagrangian smoothed particle hydrodynamics (SPH) technique, where fluid elements are modeled with particles, to treat the intergalactic gas. We compare the convergence behavior of the codes in flux statistics as well as the degree to which the codes agree in the converged limit.  We find good agreement overall with differences being less than observational uncertainties, and a particularly notable $\lesssim$1\% agreement in the 1D flux power spectrum. This agreement was achieved by applying a tessellation methodology for reconstructing the density in \CRKHACC instead of using an SPH kernel as is standard practice. We show that use of the SPH kernel can lead to significant and unnecessary biases in flux statistics; this is especially prominent at high redshifts, $z \sim 5$, as the Lyman-$\alpha$ forest mostly comes from lower-density regions which are intrinsically poorly sampled by SPH particles.
\end{abstract}




\section{Introduction}
At high redshift, the vast majority of the baryon content of the universe is present in the intergalactic medium (IGM). Neutral hydrogen in the IGM scatters light at 1216 \AA, producing characteristic absorption features in the spectra of distant quasars (QSO), dubbed the Lyman-$\alpha$ (\lya) forest. Observationally, the \lya~forest has been investigated for over half a century, but its recognition as a crucial source of cosmological information is more recent \citep{croft98, rauch_annrev}. Effectively, each quasar spectrum yields a nonlinear one-dimensional map of the IGM density at high redshifts. On large scales, the gas follows the matter density and is pressure-smoothed on small-scales; the forest traces density fluctuations in the linear to the mildly nonlinear regime at small scales, down to 0.5 h$^{-1}$Mpc, and at high redshifts, in the range $2<z<5$.

In its ability to probe smaller scales, the \lya\ forest is sensitive to the small-scale matter clustering signal -- potentially filtered by the effects of relativistic particles (e.g., massive neutrinos or exotic dark matter (DM) candidates) -- without being strongly affected by the highly nonlinear gravitational collapse and galaxy formation details that plague low-redshift probes. As such, it has been used to put strong constraints on the sum of neutrino masses~\citep{PalanqueDelabrouille2015a,PalanqueDelabrouille2015b,Yeche2017,PDB2019}, on warm dark matter (WDM) mass~\citep{viel_constraining_2005,viel_how_2007,viel_warm_2013,Baur2016,yeche_constraints_2017,Baur2017,PDB2019} and on fuzzy dark matter models~\citep{irsic_first_2017,armengaud_constraining_2017}. The \lya~ forest is also sensitive to the thermal state of the IGM and can therefore set constraints on reionization and on its thermal history~\citep{Hui1997,Zaldarriaga2001,Bolton2009,Meiksin2009,Becker2013,Lee2015,McQuinn2016,Boera2019,Walther2019,Gaikwad2021}.


The \lya~forest probes matter clustering as traced by neutral hydrogen on both large and small spatial scales. Flux transmission correlation measurements on large scales, $\sim$ 100 h$^{-1}$Mpc, are usually performed using pixels across different lines of sight (LOS) and have been used to measure the Baryon Acoustic Oscillation (BAO) peak with the Baryon Oscillation Spectroscopic Survey (BOSS) and eBOSS data at $z\sim 2.4$ \citep{Slosar2011,Slosar2013,Bautista2017,duMasdesBourboux2017,deSainteAgathe2019} providing additional constraints on dark energy. On the other hand, information on small scales is accessible through correlations along the LOS, or equivalently through the 1D power spectrum, $P_{\rm 1D}$, for which measurements can be divided into two categories. The first category consists of a large sample of medium resolution quasar spectra, in the tens of thousands, from the Sloan Digital Sky Survey (SDSS-III,~\citealt{Eisenstein2011}) BOSS program~\citep{Dawson2013} and the SDSS-IV~\citep{Blanton2017} eBOSS~\citep{Dawson2016} extension. These measurements were performed for scales  $0.001<k<0.02\,\rm~km^{-1}$s and redshifts $2.1<z<4.7$~\citep{McDonald2006,PalanqueDelabrouille2013,Chabanier2019a}, with the latest measurement being based on $>45,000$ quasar spectra and reaching $\sim 1.5\%$ statistical precision already at low redshifts. The second category of measurements originates from 8m-class telescope data sets, e.g., Keck/HIRES, VLT/UVES, or X-Shooter~\citep{Vogt1994,Dekker2000,Vernet2011}, which are significantly smaller (tens to hundreds of spectra) but with much higher resolution and higher S/N~\citep{viel_warm_2013, irsic_lyman-alpha_2017,walther_new_2018,Boera2019,Gaikwad2021,Karacayli_2022}, probing both smaller scales and extending the redshift range to $1.8<z<5.4$, albeit at lower statistical precision.

The improvement in instrumentation and data analyses of future and ongoing \lya\ surveys, such as the Dark Energy Spectroscopic Instrument (DESI, ~\citealt{Levi2013,DESI2016}) or WEAVE-QSO~\citep{Pieri2016}, will  strongly increase the precision of $P_{\rm 1D}$ observational measurements. In particular, DESI will observe a total of $\sim$ 800,000 \lya\ quasars with $z \geq 2.1$, i.e. about 4 times the largest data set to date. Also, thanks to the strong improvement in resolution -- factor of two better than eBOSS -- and the observation of fainter quasars, made possible with better target selection and the larger mirror, DESI will be able to measure $P_{\rm 1D}$ down to scales of 5~h$^{-1}$Mpc at redshifts as high as $z\sim 4.9$ at an unprecedented accuracy, 1\% at low redshifts and a few percent at the highest redshifts.

However, robust cosmological constraints using small-scale \lya\ data can only be achieved if theoretical predictions from hydrodynamical simulations can reach percent-level accuracy in order to match the quality of the observations. At high redshifts and small scales, gravitational clustering is in the mildly nonlinear regime, and because the \lya\ forest is sensitive to the intrinsic properties of the baryonic gas and details of reionization, no analytical solution for the \lya\ flux fluctuations over time is available. One must therefore resort to cosmological hydrodynamical simulations for obtaining accurate theoretical predictions.

Structure formation in the universe is often studied by gravity-only N-body methods. (For a recent review, see ~\citealt{Angulo2021}.) While these techniques do not include gas dynamics and a host of astrophysical effects, they are by now relatively well-understood. Internal tests and code comparison campaigns have shown that different N-body codes (often using different algorithms) can agree at the percent level on small scales, depending on the chosen force and mass resolution. The situation is quite different, however, for the significantly more complex case of cosmological hydrodynamical codes, where even without including subgrid physics modeling, different numerical methods have historically disagreed at the level of tens of percent~\citep{Frenk1999,Agertz2007,Sembolini2016}. As methods have been sharpened, and new ones developed, the results from different gas dynamics schemes have come closer. In particular, comparisons of (non-radiative) Nyx and \CRKHACC simulations for individual galaxy clusters have shown excellent agreement, without the need to add any unphysical conductivity or switches of various types to the CRKSPH solver~\citep{Frontiere2022}. Nevertheless, potential problems with over- and under-mixing of gas, implementation of different types of numerical viscosity schemes, differing evolution algorithms, and choices of various limiters, all contribute to making code comparison a nontrivial task.

It is common practice to test code accuracy on a number of idealized problems, where either analytical or highly accurate numerical solutions are known. These problems often possess high degrees of symmetry or reduced dimensionality, which do not make them very realistic. Codes are often tuned to address specific types of problems so that they may do well on certain of these tests and not so well on others. It is not easy, therefore, to predict how a particular method will perform in a given realistic scenario, especially since there is no rigorous a priori theory of error convergence for any method. Consequently, comparing different methods and codes (with default settings) on multiple realistic test problems is a reasonable way to assess the robustness of the numerical methods. It avoids problems with false convergence of a given algorithm; if a variety of methods, based on different assumptions and techniques, all agree well, confidence in the final result is certainly enhanced.

Although there are multiple classes of methods, the differences between smoothed particle hydrodynamics (SPH) and Eulerian techniques have been highlighted previously~(\citealt{Agertz2007,Hubber2013}). The main differences were related to SPH's inability to resolve fluid instabilities such as Kelvin-Helmholtz or Rayleigh-Taylor and the consequential effects on mixing of multi-phase media. Modern SPH methods \citep[e.g.,][]{Read2012,Saitoh2013,Hopkins2013,Frontiere2017} have been developed to overcome these difficulties and have been incorporated in cosmological hydrodynamics codes. In any case, modeling the \lya\ forest is much simpler (smooth single-phase gas, with shocks limited to high-density regions, which have little effect on the measured signals), and therefore one may expect different methods to agree better than in other, more difficult, astrophysical circumstances, although previous comparisons still reported differences at the $\sim$5-10\% level \citep{Regan2007, Bird2013, Walther2021} between different codes.  Due to advances in computational methods and computing power, as well as increased accuracy of observational data, it is of considerable interest to examine agreements between different codes at higher levels of accuracy.

In this paper, we therefore examine the current state-of-the-art in \lya\ forest statistics as reproduced by optically-thin hydrodynamical simulations. We study the convergence of \lya\ forest properties using two hydrodynamical codes following completely different numerical approaches. One is the Lagrangian, particle-based, Conservative Reproducing Kernel SPH (CRKSPH) method~\citep{Frontiere2017} implemented in the \CRKHACC code~\citep{Habib2016,Frontiere2022} and the other is the Eulerian, grid-based, finite-volume hydrodynamical method used in the Nyx code~\citep{Almgren2013, Sexton2021}. Good levels of agreement between these codes in realistic cosmological simulation studies will significantly bolster our confidence that both are converging to the correct result. Indeed, by improving the sampling of the density field in lower-density regions with the (volume-weighted) Delaunay Tesselation Field Estimator (DTFE)~(\citealt{schaap07}, \citealt{Rangel2016}) for \CRKHACC, we find that the \lya\ 1D power spectra of the two codes agree at better than 1\% at the redshifts of interest, $2.0\leq z \leq 5.0$, without any fine tuning.

The plan of this paper is as follows. In Section~\ref{sec:sims} we describe the different ways in which Nyx and \CRKHACC solve the gravitational and hydrodynamics equations along with a description of the simulation sets we perform and analyze for the purpose of the comparison. In Section~\ref{sec:methods} we outline the numerical methods used to derive the \lya\ properties from the simulation snapshots. We discuss morphological differences in Section~\ref{sec:morpho} and differences in the temperature and density fields between the two codes in Section~\ref{sec:densitytemp}. We present results on the \lya\ flux statistics, the mean flux, flux probability distribution function and (1D and 3D) flux power spectrum in Section~\ref{sec:flux} for small boxes with varying spatial resolution, and study the impact of box size in Section~\ref{sec:boxsize}. We finally conclude in Section~\ref{sec:conclusion} with a discussion of the main results.

\section{Simulations}
\label{sec:sims}

At first sight, it appears relatively straightforward to directly simulate the \lya\ forest. The gas that creates the absorption features -- the diffuse IGM -- is only at moderate overdensity, $0 \leq \delta \leq 10$, making the simulation task relatively modest in difficulty. The evolution of the gas depends on gravity and pressure forces, and some simple radiative processes. Since the gas is very close to the primordial composition, the background ionizing radiation is relatively spatially uniform, and since almost all of the gas (by volume) is optically thin to the radiation, the required simulation machinery is not complicated. At first order, any cosmological hydrodynamics code modeling dark matter and gas with uniform radiative heating and cooling should be able to adequately capture the diffuse IGM. (See~\citealt{Meiksin2009} and~\citealt{McQuinn2016} for thorough reviews of the relevant physics.) 

It is important to note, however, that there are secondary effects linked to galaxy formation and evolution, such as sources of galactic feedback, inhomogeneity of the ionizing background, or heating from helium reionization, which can strongly impact the thermal state and distribution of gas in the IGM, affecting $P_{\rm 1D}$ at the several percent level~\citep{Viel2013,Chabanier2020,MonteroCamacho2021}. Before incorporating these effects, it is important to confirm that the primary-level modeling of the \lya\ forest is in fact robust at the $\sim 1\%$ level, the main purpose of this paper.

We will employ simple -- but still fairly realistic -- assumptions regarding the physics of the \lya\ forest, which is based on a time-evolving, spatially uniform UV background. Simulations of this kind neglect the effects of inhomogeneous reionization, which produces temperature and UV background fluctuations on very large scales \citep{Onorbe2019}, and neglect galaxy formation models. 
We emphasize that this type of simulation is used as a forward model in virtually any recent inference approach applied to the \lya\ power spectrum \citep[for example, see:][]{Boera2019, Walther2019, PDB2019, Pedersen2021, Walther2021}.
In the following, we first describe the codes we use, and then present the suite of simulations performed for this work.

\subsection{Nyx}
\label{sec:nyx} 

Nyx is a publicly available\footnote{https://amrex-astro.github.io/Nyx/}, parallel, adaptive mesh, cosmological simulation code that solves the equations of compressible hydrodynamics describing the evolution of baryonic gas coupled with an N-body treatment of the dark matter in an expanding universe \citep{Almgren2013, Sexton2021}. 
Nyx's hydrodynamics is based on an Eulerian formulation, which is a very efficient approach for the low-density regions of the intergalactic medium. The code uses a second order (dimensionally-unsplit) piecewise linear (PLM) or piecewise parabolic method \citep[PPM,][]{ppm} to construct the fluxes through the interfaces of each cell. The Poisson equation for self-gravity of the gas and dark matter is solved using a geometric multigrid method.

Nyx is built on the AMReX \citep{AMReX} adaptive mesh refinement (AMR) library and is written in C++. The approach to AMR uses a nested hierarchy of logically-rectangular grids with simultaneous refinement in both space and time. AMR levels are advanced at their own timestep (sub-cycling) and jumps by factors of 2 and 4 are supported between levels. The integration algorithm on the grid hierarchy is a recursive procedure in which coarse grids are advanced in time, fine grids are advanced multiple steps to reach the same time as the coarse grids and the data at different levels are then synchronized. A separate synchronization is done for gravity.
We use MPI to distribute AMR grids across nodes and use logical tiling with OpenMP to divide a grid across threads for multi-core CPU machines (exposing coarse-grained parallelism) and/or CUDA/HIP/DPC++ to spread the work across GPU threads on GPU-based machines (fine-grained parallelism). 

 Details of Nyx's \lya\ forest modeling are given in \citealt{Lukic2015}, but we quickly summarize it here as well.  To model the \lya\ forest, Nyx follows the abundance of six species: neutral and ionized hydrogen, neutral, once and twice ionized helium, and free electrons. For these species, all relevant atomic processes -- ionization, recombination, and free-free transitions are modelled. Heating and cooling source terms are calculated using a sub-cycled approach in order to avoid running the whole code on a short, cooling timescale.  It is assumed that all gas elements are optically thin to ionizing photons, such that their ionization state can be fully described by a uniform and isotropic UV+X-ray background radiation field~\citep{Onorbe2017}. Nyx also has the capability to simulate inhomogeneous reionization models~\citep{Onorbe2019}, which affect \lya\ observables at large scales.

\subsection{CRK-HACC}
\label{sec:hacc}

\CRKHACC is the newly developed hydrodynamic extension of the high-performance cosmological N-body code \HACC \citep[Hardware/Hybrid Accelerated Cosmology Code;][]{Habib2016} which couples the gravity solver of \HACC with a Lagrangian hydro solver based on the CRKSPH method \citep{Frontiere2017}. As a modern development of SPH \citep{lucy77, gingold77}, CRKSPH leverages its key benefits (precise enforcement of conservation laws, Galilean invariance, ease of parallelization, and inherently adaptive refinement), while at the same time addressing many of the original difficulties with SPH (e.g., zeroth-order field reproduction, overly aggressive artificial viscosity models). The gravity solver in \CRKHACC uses a high-order spectral particle-mesh method handing off at small scales to short-distance algorithms (FMM, P$^3$M, tree) depending on the choice of computational architecture. The implementation of the hydro solver in \CRKHACC follows the initial design philosophy of \HACC in order to achieve easy portability and full scalability on all modern supercomputing platforms including heterogeneous CPU+GPU systems. Full details of the \CRKHACC framework are provided in \citet{Frontiere2022}. Although originally designed for large-volume, high dynamic-range cosmology simulations, \CRKHACC is easily adapted to the lower dynamic range milieu characteristic of the \lya\ forest, as described further below.

The \CRKHACC simulations performed here allow for radiative cooling and heating with the assumption that the gas remains optically thin while exposed to a spatially uniform and time-varying ultraviolet background. We assume a primordial abundance of hydrogen and helium with recombination, collisional ionization, and cooling rates taken from \citet{Lukic2015}. The energy evolution of each individual particle is evaluated from the resulting cooling function using the exact integration scheme of \citet{townsend:2009}. 
In these runs, the central densities of collapsed objects tend to grow quite large since we have not enabled any astrophysical feedback mechanisms. To avoid computational inefficiencies associated with resolving these over-cooled regions (due to its Lagrangian nature, \CRKHACC is automatically adaptive) we create collisionless star particles out of any gas particle whose density exceeds 1000 times the mean density and has a temperature below $10^5$~K (this is analogous to the {\smaller{QUICKLYA}\xspace} option in the Gadget code). We have checked that this does not impact the \lya\ measurements, which are instead sensitive to lower density regions.

There are two parameters controlling the force resolution within \CRKHACC. In the first place, the gravitational force resolution has a fixed scale that is set by the Plummer softening length, $r_{\rm soft}$. Our fiducial choice for each run here is to set the softening length to be $1/8$ the mean inter-particle separation (i.e., $r_{\rm soft} = L/8N$ where $L$ is the box size and $N$ is the one-dimensional particle count of either dark matter or baryon particles). This was used for each run except for H\_L10R78 where we set $r_{\rm soft} = L/16N$ to ensure that the softening length was
no larger than 5~$\rm h^{-1}kpc$ in any of the runs. The hydrodynamic force resolution, on the other hand, is spatially adaptive to the local gas density, as specified by the SPH smoothing lengths, $h$, of individual baryon particles. In general, the number of neighbors encompassed within a sphere of radius $h$ centered on each baryon particle is $n_{\rm nb} = (4\pi/3)n_h^3$, where $n_h$  is an adjustable parameter. The standard choice in \CRKHACC takes $n_h = 4$ meaning that each smoothing sphere contains roughly 268 neighbors.

\subsection{The Simulation Set}

We performed a set of 12 simulations, comprising of 6~\CRKHACC runs and 6~Nyx runs. Simulation box sizes of 10 and 40~$\rm h^{-1}Mpc$ were used, with the number of cells/gas particles varying from $128^3$ to $2048^3$.
A summary of the simulation characteristics is presented in Table~\ref{tab:lyasims}. We also use a set of seven additional \CRKHACC simulations with $256^3$ gas particles to test the convergence of \lya\ statistics with respect to the gravitational softening length, $r_{\rm soft}$, and the number of neighbors to define the SPH smoothing length, $n_{\rm nb}$. For the number of neighbors, we use $n_{\rm nb} = 268$ for most of the simulations. Lowering $n_{\rm nb}$ was found to provide increased contrast in the density field, but it also increases noise due to the high-order nature of the CRKSPH formalism used in \CRKHACC. We also use $n_{\rm nb} = 113$ to study the impact of the number of neighbors on the convergence of \lya\ flux statistics in low-resolution simulations. Both \CRKHACC and Nyx include radiative heating from an ultraviolet background as well as radiative cooling from a primordial composition of hydrogen and helium. We use the ``middle reionization'' ultraviolet background model from \citet{Onorbe2017}. We emphasize that, aside from the discussion presented above, no parametric tuning was attempted to bring the codes into better agreement for the comparison runs.

We generated initial conditions for a $\Lambda$CDM cosmology consistent with the WMAP-7 \citep{Komatsu:2011} measurements. In particular, we have $\Omega_b = 0.045$, $\Omega_m = 0.27$, $\Omega_\Lambda = 0.73$, $h = 0.70$, $\sigma_8 = 0.8$ and $n_s = 0.95$. Initial conditions were generated at a starting redshift of $z = 99$ using the Zel'dovich approximation \citep{Zeldovich1970} to perturb $2\times N^3$ dark matter plus baryon particles each arranged on a uniform mesh that are offset from each other by half the mesh spacing. The Zel'dovich displacement field was computed on a $2048^3$ ($4096^3$) mesh for the $10\ (40)~\rm h^{-1}{\rm Mpc}$ box allowing particle counts up to $N = 1024\ (2048)$ without requiring any interpolation from the displacement field. In this way, all runs at fixed box size pertain to the same random realization of the universe. The simulations were run to a final redshift of $z = 2$ (except for the H\_L10R10 and H\_L40R20 runs which were stopped at $z = 3$ due to the computational cost of simulating further) and we focus on \lya\ measurements in the range $2 \leq z \leq 5$ relevant for both high and medium resolution surveys.

  \begin{table*}
\centering
\begin{tabular}{ccccc}
\hline
Name & Box sixe [Mpc/h] & Gas cells/particles & Average resolution [kpc/h] \\
\hline
L10R78 & 10 & $128^3$ & 78  \\
L10R39 &  10 & $256^3$ & 39  \\
L10NR20 &  10 &  $512^3$ & 20  \\
L10R10 &  10 & $1024^3$ & 10  \\
L40R39 &  40 &  $1024^3$ & 39  \\
L40R20 &  40 & $2048^3$ & 20  \\
\hline
\end{tabular}
\label{tab:lyasims}
\caption{The set of simulations with the name of the simulation, the box size in $\rm h^{-1}Mpc$, the number of gas cells/particles and the equivalent average resolution in $\rm h^{-1}kpc$, which is the exact resolution in the Nyx case and the mean inter-particle separation in the \CRKHACC case. When referring to Nyx and \CRKHACC outputs we write N\_ and H\_ respectively, followed by the simulation label.}
\centering
\end{table*}


\section{Numerical methods}
\label{sec:methods}
In this section we describe the simulation output data and how it is processed to produce synthetic \lya\ forest sightlines. For each simulation, we have snapshots at $z =$ 2, 3, 4 and 5, except for H\_L10R10 and H\_L40R20 for which we do not include $z = 2.0$. For most of the comparisons, we include only redshifts $z=3$ and $z=5$, which are relevant for medium and high resolution observational data sets, respectively.

The first step to compute \lya\ forest quantities is to extract sightlines for each simulation box. For the Nyx outputs, we take skewers along the three simulation axes, keeping periodic boundary conditions with the rays passing through all cell centers. For \CRKHACC outputs, we deposit particle quantities on a grid and follow the same procedure. In general, there are a number of ways in which particle data can be interpolated to a grid. We choose here to use an SPH scattering method that interpolates particle quantities onto cell centers via an SPH kernel. This is the most natural choice to create a volume-filling interpolation that makes use of the smoothing lengths internally computed within \CRKHACC. In contrast, methods such as nearest grid point (NGP) or cloud-in-cell (CIC) interpolation will suffer from empty cells particularly in low-density regions that are important for the \lya\ calculation. 

The SPH interpolation works in the following manner: for each particle quantity $\psi$, we compute the field $\psi(x_i)$ at each mesh point ${\bf x}_i$ via
\begin{equation}
\psi({\bf x}_i) = \frac{\sum_j \psi_j W_{ij} V_j}{\sum_j W_{ij}V_j},
\label{eq:psii}
\end{equation}
where $V_j = m_j/\rho_j$ is the volume of a particle with mass $m_j$ and density $\rho_j$ while $W_{ij} = W({\bf x}_i - {\bf x}_j,h_j)$ is the SPH weight of a particle at location ${\bf x}_j$ with smoothing length $h_j$. The sums in Eqn.~(\ref{eq:psii}) are performed over all $j$ particles whose smoothing spheres intersect with the mesh point ${\bf x}_i$ while the functional form of $W_{ij}$ corresponds to a 4th-order ($C^4$) Wendland function \citep{Dehnen2012}. The particle quantities that are interpolated in this manner include density, temperature, and velocity. Each of these quantities are calculated internally within the CRKSPH formalism of \CRKHACC along with the smoothing lengths of each individual particle. We find that for good convergence of \lya\ quantities we need a grid with $N_{\rm grid} = 2 \times N_{\rm part}$, where $N_{\rm grid}$ is the number of grid elements on each side and $N_{\rm part}$ is the number of particles on each side.

After initial comparisons of the \CRKHACC and Nyx results, we found that the main differences in their \lya\ measurements could be attributed to differences in the density field. Visual inspection of sightlines showed that the \CRKHACC density field was noticeably smoother and slower to converge with increasing resolution compared to Nyx. Interestingly, this was only observed in the density field whereas temperature and velocity displayed stronger agreement with Nyx. The contrast in the density field could be enhanced by lowering the SPH neighbor count, $n_{\rm nb}$, used in \CRKHACC. However, making this number too small ($n_h <= 2$ or $n_{\rm nb} \lesssim 34$) resulted in numerical issues as the high-order CRKSPH solver is sensitive to low particle sampling noise. 

To investigate the discrepancy in density fields further, we carried out separate \CRKHACC \lya\ measurements with a density field computed using a DTFE method, which is known to have good properties for density estimation at low particle densities, limiting the effects of shot noise. In this case, a parallel DTFE code~\citep{Rangel2016} developed specifically for cosmological applications was used to compute a density for each particle that is subsequently interpolated to the grid via Eqn.~(\ref{eq:psii}), and led to much improved results, as discussed further below. In this way, we are replacing only the SPH density estimate with a DTFE density estimate while preserving the use of an SPH interpolation kernel to map the particle density to the grid. We also tried replacing the SPH grid interpolation with a DTFE grid interpolation, but found that this only served to pick up small-scale noise in the DTFE density field along the gridded sightlines. In other words, the advantage of the DTFE is to provide a better particle density estimate in low-density regions while not necessarily providing a better particle to grid interpolation scheme. In what follows, we use the notation \CRKHACC.sph and \CRKHACC.dtfe to distinguish between \lya\ measurements derived from density fields obtained using the SPH density taken from the simulation output and from the DTFE, respectively. Note that both cases use the same temperature and velocity fields.

The second step is to compute the (normalized) transmitted flux $F$ at every pixel, with $F = e^{-\tau}$ where $\tau$ is the optical depth for \lya\ photon scattering. The latter is defined as
\begin{equation}
\tau_\nu = \frac{\pi e^2}{m_e c}f_{12}\int \frac{n_{\rm HI}}{\Delta_{\nu_{D}}}\phi_\nu dr,
\end{equation}
where $\nu$ is the observed frequency, $e$ is the electron charge,  $c$ is the speed of light, $f_{12}$ is the oscillator strength for the \lya\ resonance transition, $n_{H_I}$ is the neutral hydrogen density, $\Delta \nu_{D} = (b_T/c)\nu_0 = (\sqrt{2 k_B T / m_H} / c)\nu_0 $ is the Doppler width with $b_T$ the Doppler parameter, $m_H$ the mass of hydrogen, $k_B$ the Boltzmann constant and $\phi_\nu$ is the line profile. 

In general, the line profile is a Voigt profile, but we use the Doppler profile instead for several reasons. For line center optical depths of less than 1000, the Doppler profile gives identical fluxes. We are only interested in \lya\ forest systems, which have line center optical depths less than 10. For Lyman Limit Systems (LLS) and Damped \lya\ systems (DLAs), our simulations are not designed to produce the correct density and temperature for HI density mapping in any case. The HI density in these systems should have self-shielding corrections, which cannot be evaluated properly without coupled radiative transfer-hydrodynamics in the simulations. If we were to use Voigt profiles with these high column density systems, the damping wings would not only be inaccurate, but those errors would then contaminate nearby regions. The Doppler profile is
\begin{equation}
\phi_\nu = \frac{1}{\sqrt{\pi}} \exp \left [ -\left( \frac{\nu - (1-\frac{\nu_{\rm ||}}{c})\nu_0}{\Delta \nu_D} \right)^2 \right].
\end{equation}
In velocity space, peculiar velocities modify the optical depth by shifting the absorption positions and broadening the lines. Thus, in redshift space we have
\begin{equation}
    \tau_v = \frac{\pi e^2 f_{12}\lambda_0}{m_e c H}\int \frac{n_{\rm HI}}{v_{\rm th}} \exp\left[ - \left( \frac{v-v'-v_{\rm ||}}{v_{\rm th}}, \right)^2\right] dv'
\end{equation}
where $\lambda_0$ is the rest-frame wavelength and $H$ is the Hubble expansion rate at the given redshift. All the \lya\ fields computation, along with the different flux statistics presented in the following section are performed using the gimlet postprocessing software (see for instance \citealt{Briesen2016}).

\section{Morphological comparison}
\label{sec:morpho}

In this section, we compare morphological differences between the most resolved outputs from the L10R10 runs. This comparison is only qualitative in order to highlight the differences in the properties of the simulation data between the codes. It  aims at building our understanding of the impact of these two different numerical methods on observable \lya\ flux statistics, which are studied in the following sections. 

\begin{figure*}
\centering
\includegraphics[width=0.48\textwidth]{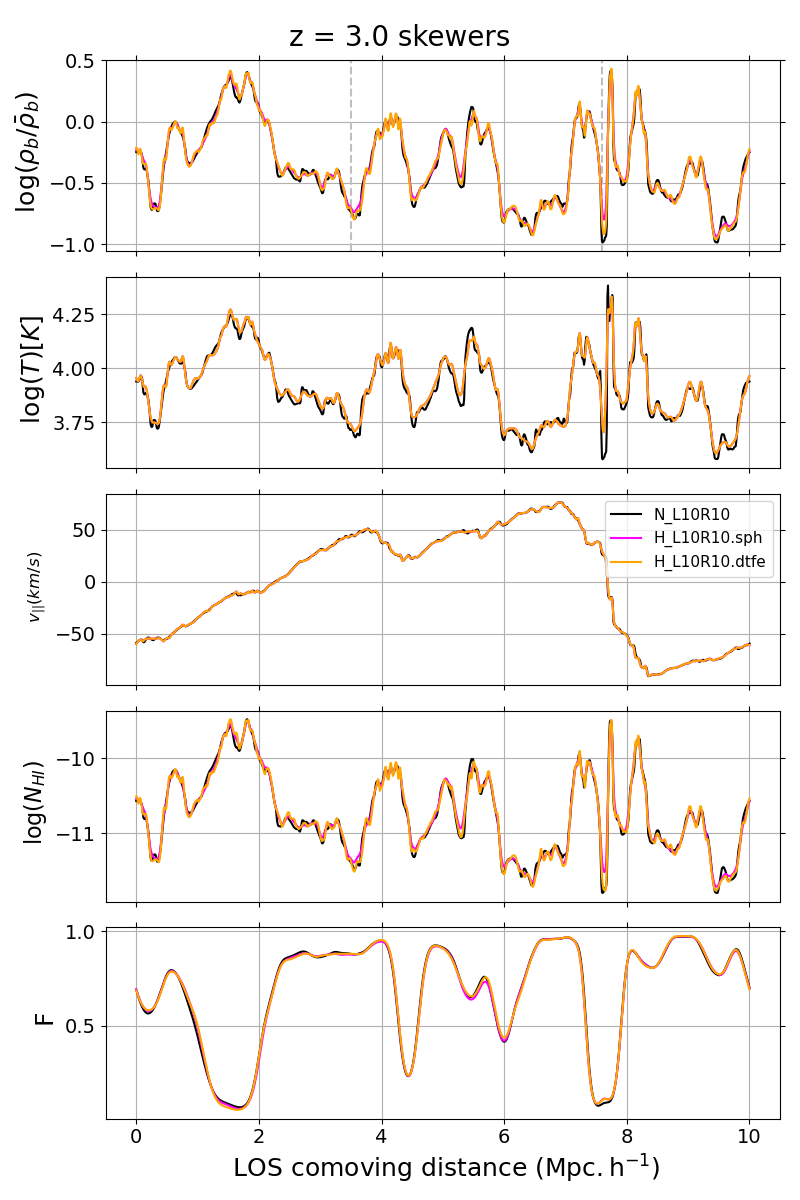}
\includegraphics[width=0.48\textwidth]{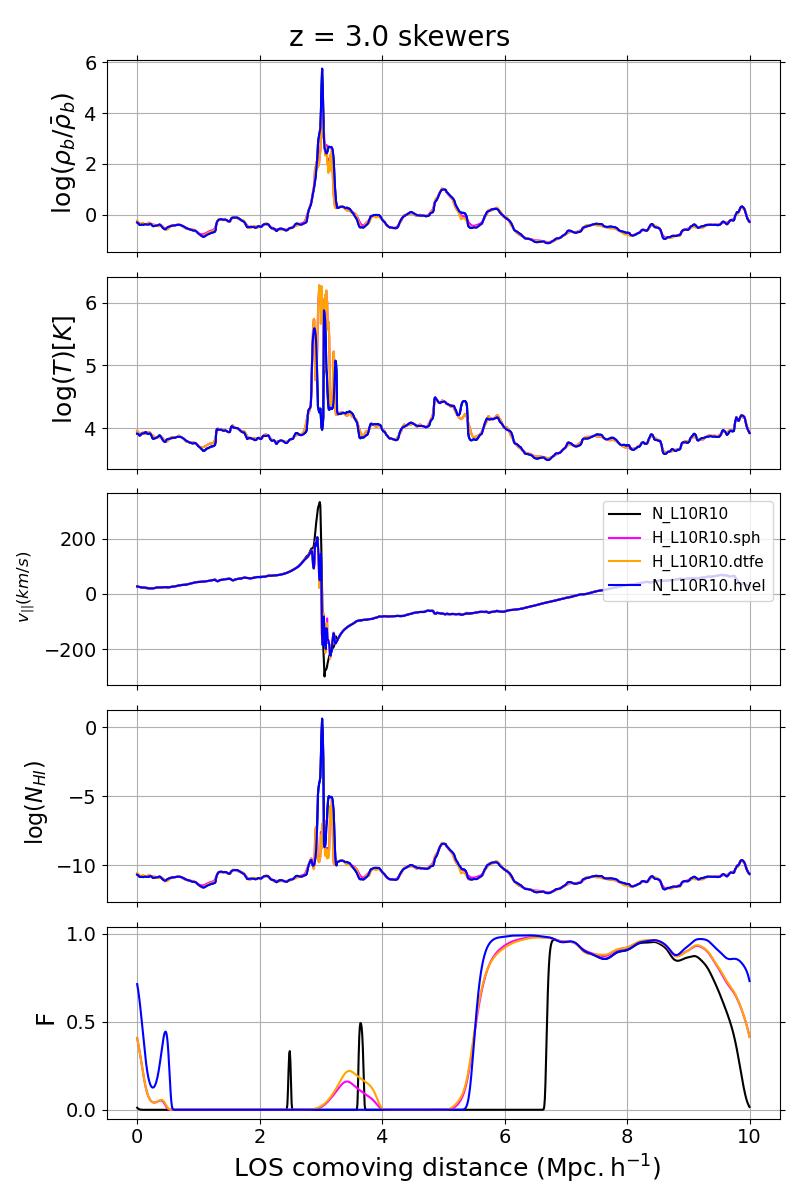}
\caption{Skewers from the snapshots in 10 $\rm h^{-1}Mpc$ with 1024$^3$ cells/particles (L10R10) at $z = 3$ for Nyx in black and  \CRKHACC using the SPH and DTFE methods to estimate the density fields in magenta and orange respectively. The baryon density $\rho_b$ (first row), the temperature $T$ (second row), the velocity parallel to the LOS (third row), the neutral Hydrogen density $n_{\rm HI}$ (fourth row) and the redshift-space flux $F$ (fifth row) are displayed. The left panel skewer goes through a low density region (for a more obvious visual  difference between the SPH and DTFE results, see the lower-resolution comparison data shown in Fig.~\ref{fig:skewer_lowres} in the Appendix), whereas the right panel skewer samples a high density region. The blue line shows the Nyx output using the \CRKHACC velocity field to compute the flux (the only panels affected are for the velocity and the flux, for further discussion, see text).}
\label{fig:skewers}
\end{figure*}

In Fig.~\ref{fig:skewers}, we show two example skewers of \CRKHACC.dtfe, \CRKHACC.sph and Nyx at $z=3$, with the baryon density, $\rho_b$, the temperature, $T$, and the redshift space flux $F=e^{-\tau}$. The skewer on the left pierces low to moderate densities with $-1.0 \leq \log(\rho_b / \bar{\rho_b}) \leq 0.5$, i.e. the typical density regime from which originates the \lya\ signal. The skewer on the right pierces a high density region with $\rho_b / \bar{\rho_b} \geq 10^{5}$. For \CRKHACC skewers, we used a grid with 2 times more cells than particles in each dimension. The low density skewer displays good agreement between Nyx and \CRKHACC. Although the initial conditions are the same, there will be differences in how structure will develop in the codes, as a function of position at low redshift. We observe two systematic discrepancies between Nyx and \CRKHACC.sph. First, the \CRKHACC.sph results tend to overestimate the density in under-dense regions, e.g., at $x \sim 3.5 \ \rm h^{-1}Mpc$ or $x \sim 7.3 \ \rm h^{-1}Mpc$ (indicated by dashed grey vertical lines), the Nyx voids are systematically deeper than those in \CRKHACC.sph. Second, while the \CRKHACC.sph and Nyx skewers capture the same overall structures, the former misses some small-scale fluctuations because of the inherent smoothing of the density estimate with the chosen parameter choices. Using the DTFE to estimate the density field of \CRKHACC significantly improves the agreement between the two codes in the very low density regions (with voids being deeper) but still tends to over-smooth small-scale fluctuations. The density fields from the same skewer in lower resolution runs can be found in App.~\ref{sec:lowres}, where typical differences between the two codes are easier to catch by eye than on the high-resolution outputs.   

In the right skewer of Fig.~\ref{fig:skewers}, we intercept a high-density spike, around $x \sim 3.5\ \rm h^{-1}Mpc$, affected by the formation of collisionless star particles. This produces features in the density, temperature, and velocity fields. Combined with the high temperature and large velocity gradient, the disrupted region distorts the spectrum from $\sim 1$ to $\sim 6 \ \rm h^{-1}Mpc$. In particular, around 5~h$^{-1}$Mpc, we observe a 1~$\rm h^{-1}Mpc$ shift between Nyx and \CRKHACC, where the redshift-space flux goes from saturated to unsaturated. Using the \CRKHACC velocity instead of the Nyx velocity produces flux fields in much better agreement, without the 1~$\rm h^{-1}Mpc$ shift (as shown by the blue curve in Fig.~\ref{fig:skewers}). We wish to highlight this contradiction of the standard lore that high density regions do not materially affect the forest. The absorption is saturated near the position of the density spikes, but the differences in the hydrogen column density and temperature give different line shapes in the spectra. In addition, high density regions typically have large velocity gradients, meaning that any difference is spread over much larger scales in redshift space. However, these high-density peaks are so rare that they do not affect the total \lya\ power spectrum, as we will see in Section~\ref{sec:flux}.

\begin{figure*}
\centering
\includegraphics[width=1.00\textwidth]{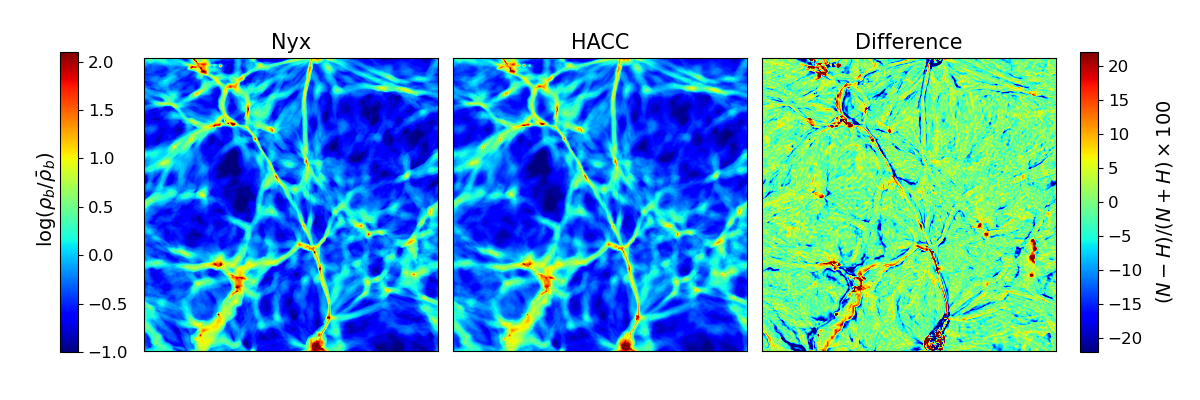}
\includegraphics[width=1.00\textwidth]{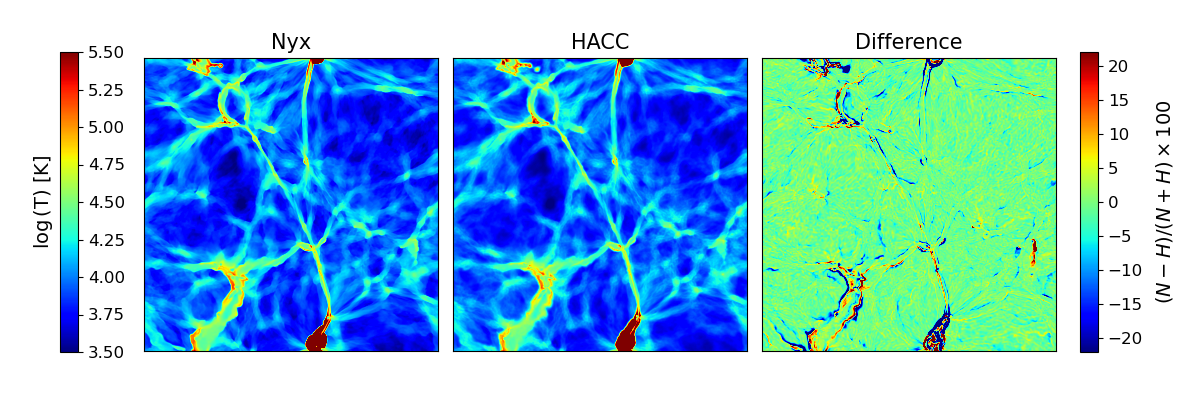}
\includegraphics[width=1.00\textwidth]{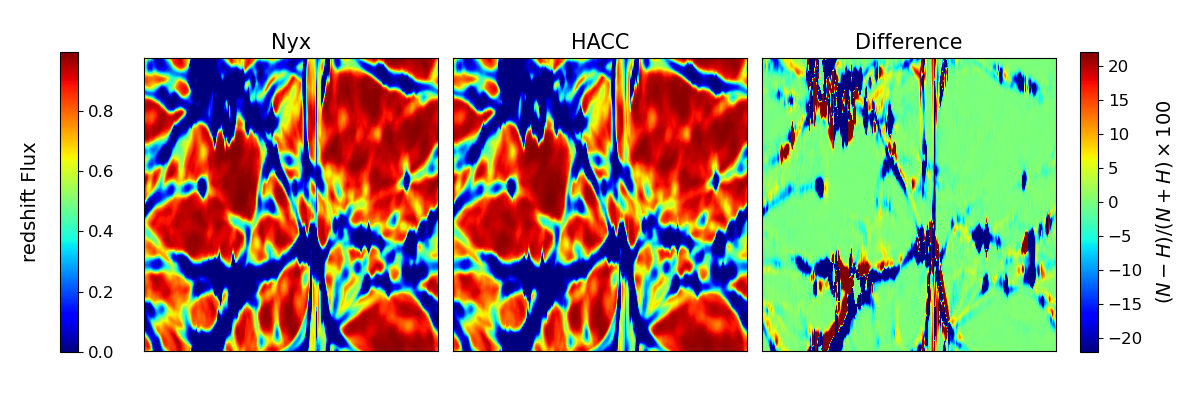}
\caption{Slices through  snapshots of the 10 $\rm h^{-1}Mpc$ with 1024$^3$ cells/particles simulations (L10R10) at $z = 3$ of Nyx (left) and \CRKHACC (middle) and the difference (right) at $z = 3$, showing the baryon density $\rho_b$ (top), the temperature $T$ (middle) and the transmitted flux $F$ (bottom). The \CRKHACC density was estimated with the SPH method.}
\label{fig:slices}
\end{figure*}

Fig.~\ref{fig:slices} shows an example slice through our simulations. We do not show slices for the DTFE \CRKHACC density field as the main differences have already been discussed. We show the baryon density and temperature and the transmitted flux from the L10R10 simulations at $z = 3$. We also include differences of the simulated fields. The LOS axis is plotted vertically, so in the  flux images, one can imagine each column of pixels as a separate spectrum. Again, the qualitative agreement is very good and we see the same structures in all panels. The major differences in density arise near the high-density regions, near the bottom and top clusters and along the filaments. The highest density regions are more extended in \CRKHACC due to the formation of star particles out of high-density gas particles. Nyx, on the other hand, allows baryons to cool and collapse much more, producing more concentrated halo density profiles and thinner filaments, as seen by the blue stripes along filaments. We observe similar patterns in the temperature maps, with the largest differences mostly in accretion shocks enveloping the filaments and halos, where the shocks in \CRKHACC are slightly puffier. Structures appear very similar in the other panels, but in flux, they are distorted in position along the LOS by peculiar velocities and broadened along the LOS by thermal broadening. Where the other panels also show smooth transitions between extreme values, the flux often has sharp features along the LOS, quickly changing from a saturated region (red) to a transmissive region (blue). The largest differences in flux once again come from the high density regions, where different structures are created by the formation of star particles in \CRKHACC and overcooling in Nyx. These largest differences arise in saturated or almost saturated regions so that they do not strongly impact the \lya\ forest. However, subtler differences arise in the low-density regions, which more significantly impact \lya\ flux statistics, as we will discuss in Section~\ref{sec:flux}.

\section{Density and temperature statistics}
\label{sec:densitytemp}
\begin{figure*}
\centering
\includegraphics[width=1.00\textwidth]{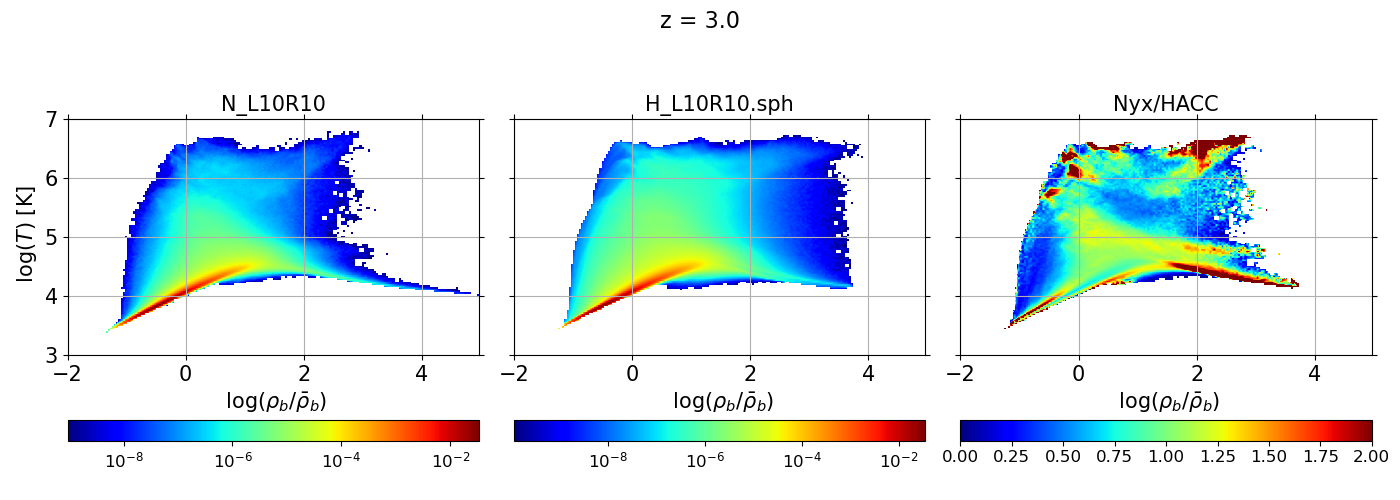}
\includegraphics[width=1.00\textwidth]{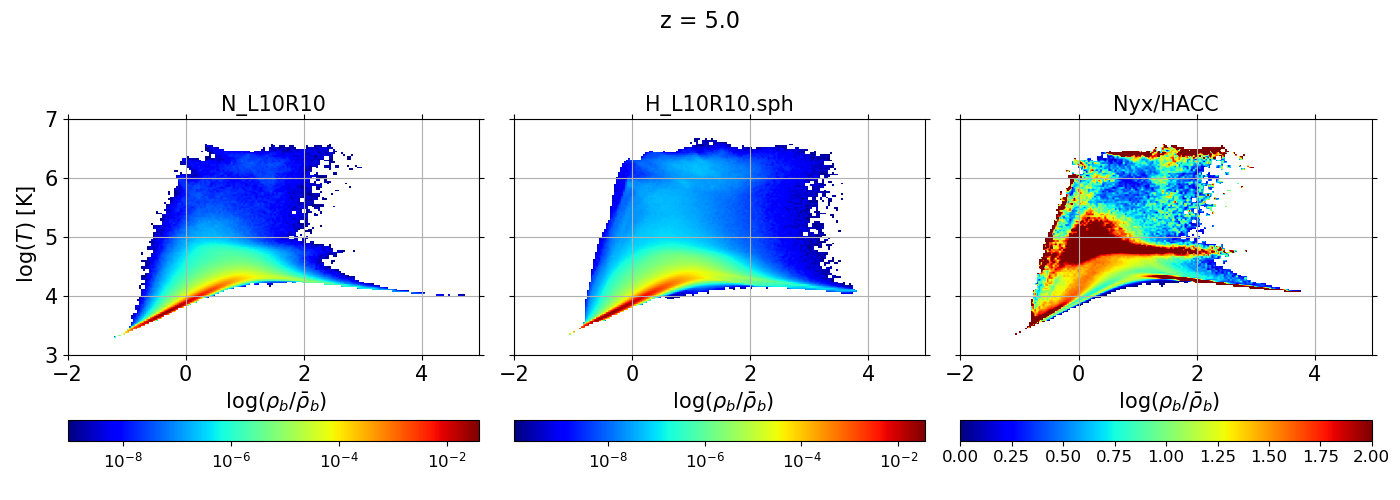}
\caption{Temperature-density diagrams for Nyx (left) and \CRKHACC (middle) outputs in a 10~$\rm h^{-1}Mpc$ box with 1024$^3$ cells/particles (L10R10) and their ratio (right). We show results at $z=3$ (top) and $z=5$ (bottom). Here we use the SPH method to estimate the \CRKHACC density field. The high gas density tail is not seen in the \CRKHACC simulation since it has been converted into stars.
}
\label{fig:TD_diag}
\end{figure*}

\begin{figure}
\centering
\includegraphics[width=1.00\columnwidth]{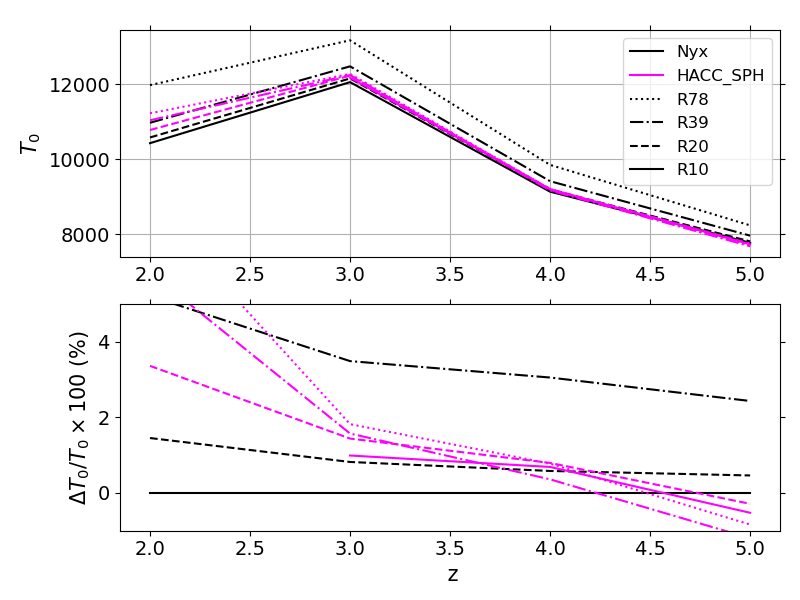}
\includegraphics[width=1.00\columnwidth]{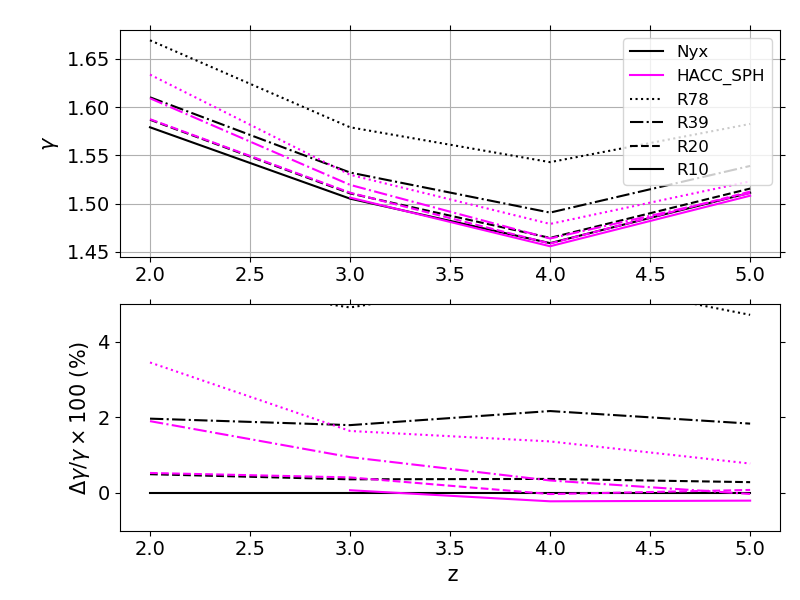}
\caption{$T_0$ and $\gamma$ evolution with redshift for all 10 $\rm h^{-1}Mpc$ simulations. Bottom panels show relative differences with respect to N\_L10R10. For \CRKHACC we use the SPH method to estimate the density field.
}
\label{fig:t0gamma}
\end{figure}

We now consider the density and temperature statistics in the simulations. Fig.~\ref{fig:TD_diag} shows the temperature density diagrams of the baryonic component in the most resolved simulations, N\_L10R10 and H\_L10R10, at $z=3$ and $z=5$, which illustrate the phases of the IGM as captured in both codes. We do not use results for the DTFE snapshots in order to ensure that we show a self-consistent relationship between temperature and density values computed internally within the simulation. The PDF is calculated as the volume-weighted histogram of $\log \rho_b$ and $\log T$. We note that the \CRKHACC particle data is not used directly, as this would result in different spatial sampling. Instead, we use the SPH values deposited on the uniform grid. We also fit the  $\rho_b$-T relation in the diffuse IGM, which is a key component of the robustness of \lya\ forest predictions. It is approximated by the power law
\begin{equation}
    T = T_0 \left(\frac{\rho_b}{\bar{\rho}_b}\right)^{\gamma-1},
\end{equation}
\label{eq:tempdensity}
where $T_0$ is the temperature at the mean density and $\gamma$ is the slope of the power law, both of which are set by the ionizing background. The fit is performed using cells in the low-density regime, with $10^{-1}\bar{\rho_b}\leq \rho_b \leq 10\bar{\rho_b}$ and $T \leq 10^5 \rm \ K$. We show the evolution of $T_0$ and $\gamma$ with redshift for all the simulations in 10~$\rm h^{-1}Mpc$ boxes in Fig.~\ref{fig:t0gamma}.

Returning to Fig.~\ref{fig:TD_diag}, we note that the diffuse phase is similar between the two codes, as expected. The $\rho_b$-T relation overlaps at all redshifts for converged simulations, although the temperature in \CRKHACC might be slightly higher at low redshift, and slightly lower at high redshift. The activation of star formation at high density in \CRKHACC is clearly apparent in Fig.~\ref{fig:TD_diag}, since points past $\rho_b \sim 10^4 \bar \rho_b$ on the galaxy phase tail are systematically lost. Finally, \CRKHACC is much faster to converge on $T_0$ and $\gamma$, in particular at high redshift where there is very little improvement with spatial resolution.

\section{Flux statistics}
\label{sec:flux}
In this section, we compare the observationally relevant \lya\ flux statistics that are commonly used to constrain astrophysics and cosmology, the 0-, 1-, and 2-point flux statistics, i.e., the mean flux, the flux probability distribution function (PDF) and the flux power spectrum. We show the comparison in the small box at $z=3$ and $z=5$, which are the most relevant for observations. We will discuss box-size effects in the following section.

\subsection{Mean flux}
\label{sec:meanflux}

\begin{figure}
\centering
\includegraphics[width=1.00\columnwidth]{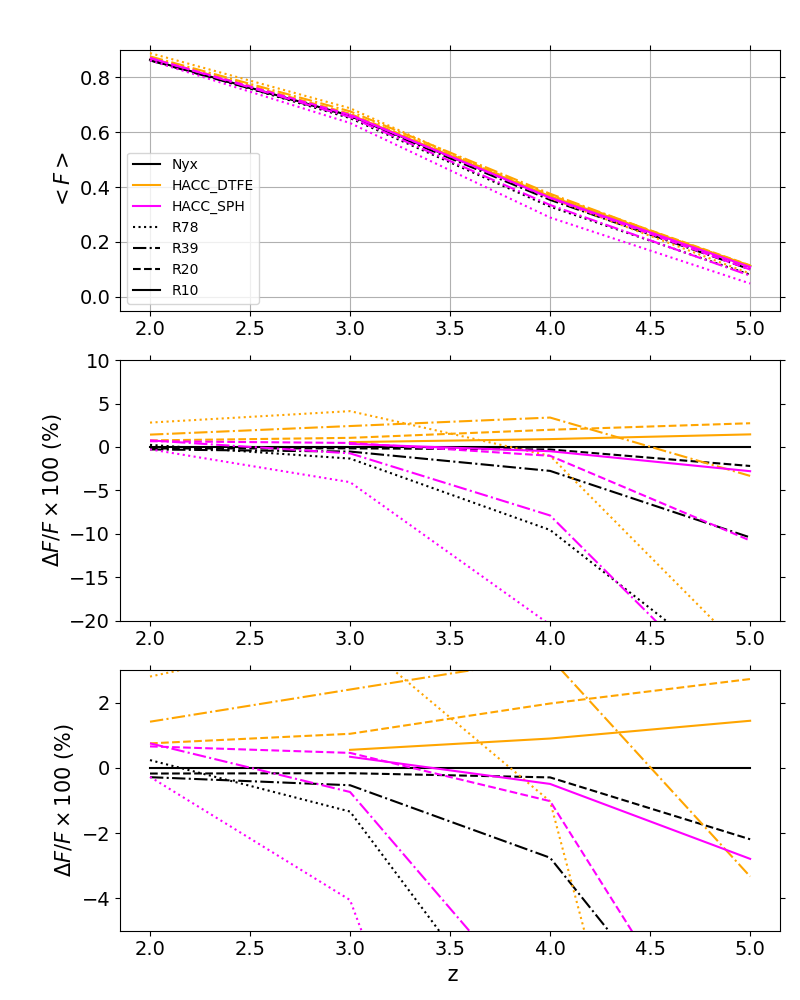}
\caption{Mean flux as a function of redshift for all Nyx and \CRKHACC simulations in 10 $\rm h^{-1}Mpc$ boxes for different spatial resolution (top). Middle and bottom panels show relative differences with respect to N\_L10R10, the most-resolved Nyx simulation with 10~$\rm h^{-1}kpc$ resolution. }
\label{fig:meanflux}
\end{figure}

The simplest \lya\ flux statistic is the mean transmitted flux, $\left< F \right>$, or equivalently the effective optical depth $\tau_{\rm eff} = -\log\left(\left< F\right>\right)$, which is a measure of the opacity in the IGM. Observations shows that it smoothly evolves from low to high values with decreasing redshift as the expansion  gradually lowers the proper hydrogen density and the UVB intensity decreases the neutral hydrogen fraction. Current mean flux measurements vary between different groups, e.g. at $z = 2.8$ $\bar{F} \sim 0.69$ at the 2-3\% level~\citep{PalanqueDelabrouille2015b,Walther2019} and $\bar{F} \sim 0.72$ at the 1\% level~\citep{Becker2007,Becker2013}. We compute the mean flux using all pixels in the box. Fig.~\ref{fig:meanflux} shows the mean flux as a function of redshift for all 10~$\rm h^{-1}Mpc$ boxes at all snapshots, varying spatial resolution only.

The two codes have a similar convergence trend, with higher discrepancies at higher redshifts. Low-resolution runs have more absorption than the high resolution ones for Nyx and \CRKHACC.sph as they overestimate the density in low-density regions. But we observe a non-monotonic convergence for \CRKHACC.dtfe at high redshift and an opposed sense of convergence at low redshifts. For low redshifts, $2 \leq z \leq 3$, both Nyx and \CRKHACC converge at the percent level, with differences between N\_L10R10 - N\_L10R20 and H\_L10R10 - H\_L10R20 below 1\%, using either SPH or DTFE for the density field.  At high redshifts, $4 \leq z \leq 5$, the mean flux is less converged with respect to spatial resolution, but still within observational uncertainties with 8\%, 1\% and 2\% differences for \CRKHACC.sph, \CRKHACC.dtfe and Nyx respectively. The increase in the spatial resolution requirement with redshift to model the \lya\ forest is fully expected because of three effects. First, the comoving filtering scale increases with redshift. Then, the majority of the \lya\ signal comes from mildly-dense regions at $z=2.0$ and from under-dense regions at $z=5$, which are more difficult to capture~\citep{Bolton2009}. And finally, thermal broadening is weaker at higher redshifts. 

\begin{figure}
\centering
\includegraphics[width=1.0\columnwidth]{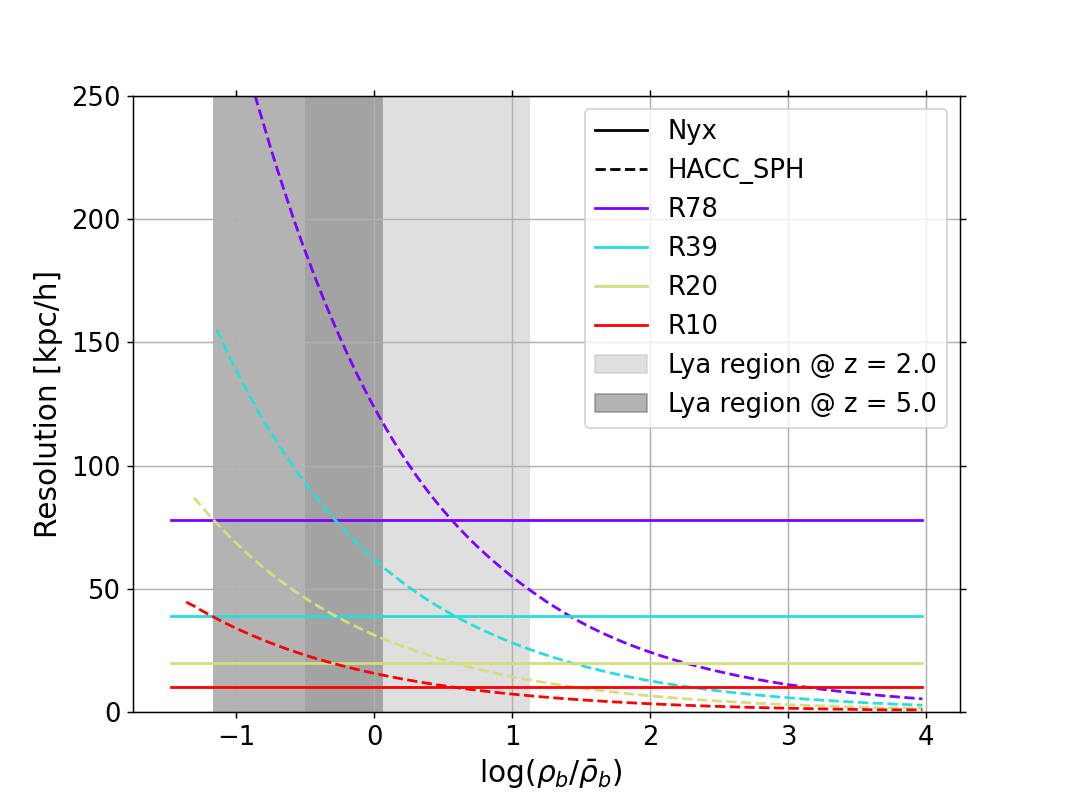}
\caption{Effective resolution as a function of baryon density for Nyx (solid) and \CRKHACC.sph (dashed) runs in 10~$\rm h^{-1}Mpc$ boxes varying spatial resolution. The light gray and dark gray shaded regions show the density regime from which originates the \lya\ forest at $z=5$ and $z = 2$ respectively.}
\label{fig:reso}
\end{figure}

An important finding is that the Nyx results converge faster than \CRKHACC.sph, as the former requires a lower average resolution than the latter to reach convergence. In Fig.~\ref{fig:reso} we show the mean effective resolution as a function of density for the two codes when varying the number of cells/particles. For Nyx, given that all cells have the same size, we simply take the resolution $r$ as
\begin{equation}
    \rm r = \frac{L}{N_{\rm cells}},
\end{equation}
where $L$ is the box size in $\rm h^{-1}kpc$ and $N_{\rm cells}$ the number of cells in one dimension. For \CRKHACC we define the effective resolution as the local inter-particle separation which we derive from the normalized smoothing length, $h/n_h$. We then bin particles in density and plot the mean normalized smoothing length for each bin in  Fig.~\ref{fig:reso}.

It is clear that  for the same number of particles/cells, if the average resolution is the same, the effective resolution in low-density regions is better for Nyx, while it is better in high-density regions for \CRKHACC. This does not come as a surprise, as particles are attracted to high-density regions at the expense of low-density regions. Therefore, a Lagrangian code will always need more particles than cells for an Eulerian code to reach the same effective resolution in the \lya\ forest. We also highlight the density regions from which the \lya\ signal originates at $z = 2$ and $z = 5$ (the lighter and darker gray bars in Fig.~\ref{fig:reso}). To define the minimum and maximum density of these regions, we consider cells with flux $0.85 \leq F \leq 0.95$ and $0.1 \leq F \leq 0.2$ respectively, and take the median density. This comparison of effective resolution with density nicely shows why Nyx and \CRKHACC are closer at low redshift compared to high redshift. The use of DTFE significantly improves the estimation of the density field in the lower density regions, making the internal convergence of the mean flux of \CRKHACC as fast as the convergence of Nyx. We also find that the improvement when using DTFE increases with redshift because of the increased importance of lower density regions to the \lya\ signal at $z=5$ than at $z=2.0$.

Equally importantly, when comparing converged results (in the sense of being ``internally converged'', i.e., when increasing the spatial resolution does not induce any further changes), the Nyx and two \CRKHACC methods each converge to the same answer with differences below 1\% for $2 \leq z \leq 3$ and 2.5\% for $4 \leq z \leq 5$. The behavior of the mean flux evolution, however, is a little different between the two codes when using the DTFE density field. At the best resolution, Nyx starts with a higher mean flux at high redshift and ends up with a lower mean flux at low redshift than \CRKHACC. We note that the number of neighbors parameter, $n_{\rm nb}$, for the \CRKHACC runs has no impact on the mean flux.

It is common practice to rescale the effective optical depth so that the simulated mean flux matches the observed mean flux, which is roughly equivalent to adjusting the intensity of the specific UVB used in the simulation. Indeed simulations of the IGM must assume a particular ionizing background and common prescriptions of the ionizing background  provide a table of the photoionization rate and heating per atom for the primordial species HI, HeI, and HeII at many redshifts. The reality is that our constraints on the ionizing background are much weaker than our knowledge of cosmology but this uncertainty can be parametrized by rescaling the simulated optical depth \citep{Lukic2015}, which is more or less equivalent to changing the photoionization rate $\Gamma$ in the simulation since $\Gamma \propto n_{\rm HI} \propto 1 / \Gamma_{\rm HI}$.
For relative flux fluctuations $\delta_F = F/ \left< F\right> -1 $, a small difference in mean flux directly propagates as an amplitude shift in power spectra. For all other flux statistics, we rescale the optical depths by a small amount to match N\_L10R10, i.e. $\bar{F} =  0.66051489$ at $z=3$ and $\bar{F} = 0.11177497$ at $z=5$.

\subsection{Flux PDF}

\begin{figure*}
\centering
\includegraphics[width=0.49\textwidth]{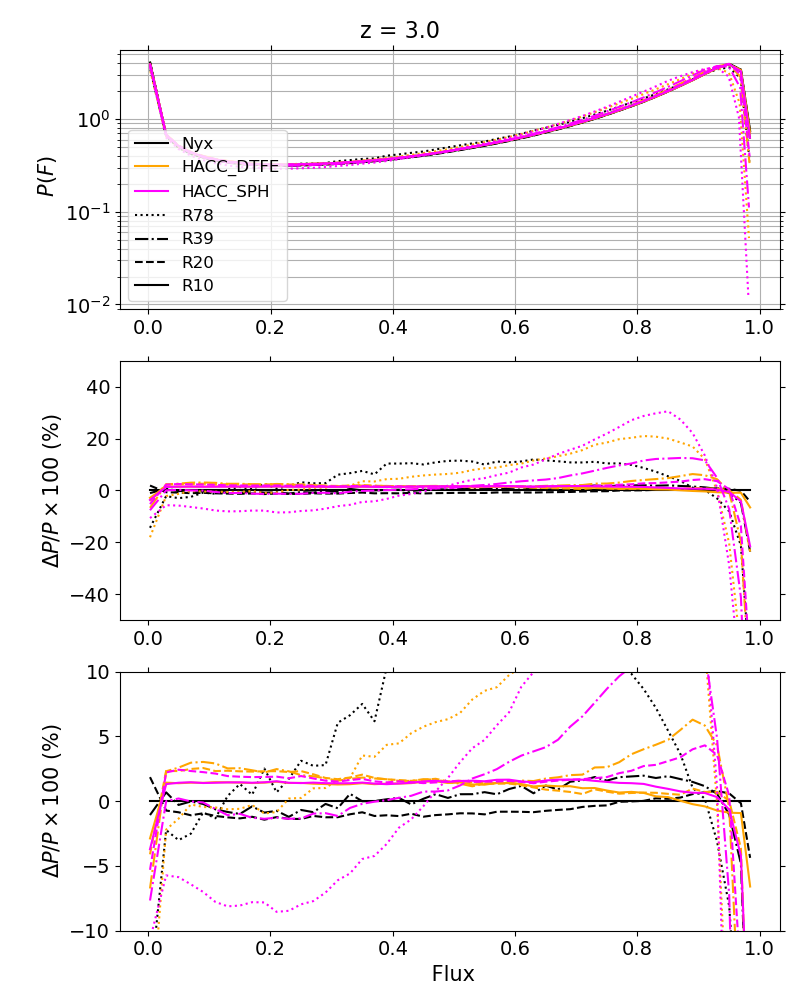}
\includegraphics[width=0.49\textwidth]{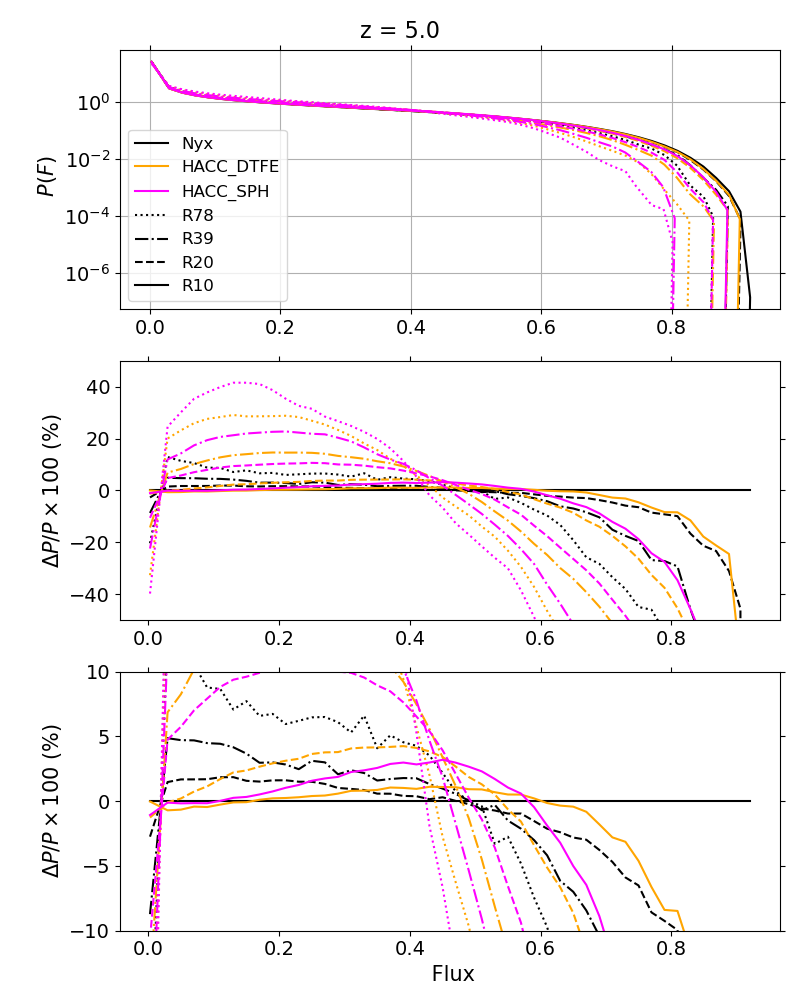}
\caption{Flux PDF for Nyx and \CRKHACC at $z=3$ (left) and $z=5$ (right). Middle and bottom panels show relative differences with respect to N\_L10R10, the most resolved Nyx simulation with 10 $\rm h^{-1}kpc$ resolution.}
\label{fig:pdf}
\end{figure*}

The flux PDF, the probability density function of the pixel fluxes, has been a probe of interest for a long time, estimated with both medium-resolution~\citep{Desjacques2007,Lee2015} and high-resolution quasar spectra~\citep{McDonald2000,Lidz2005,Kim2007,Calura2012,Rollinde2013,Rorai2017,Gaikwad2021} with precision between 5-10\%, to measure the amplitude of matter fluctuations or the thermal history of the IGM through $T_0$ and $\gamma$~\citep{Bolton2008,Viel2009,Calura2012,Lee2015,Rorai2017,Gaikwad2021}.
We compute the flux PDF, $P(F)$, in 50 equally spaced flux bins with the integral constraint $\int P(F)\rm dF = 1$. The results are shown in Fig.~\ref{fig:pdf} at $z=3$ and $z=5$.

We observe that in this case the resolution requirement also increases with redshift. At $z=3$, the convergence is relatively rapid for both codes and they agree with each other at the 1\% level except for $F \sim 0$ and $F \sim 1$. Nyx has a larger probability at the extremas and the integral constraint pushes the intermediate probabilities down relative to \CRKHACC values.  At $z = 5$, we observe a threshold in the flux above which the simulation cannot produce the flux value, since the low-density gas is missed in the low-resolution runs. This threshold is systematically higher for Nyx for the same number of cells/particles, for the resolution-related reasons detailed in Sec.~\ref{sec:meanflux}. For low-medium flux values, the results internally converge at the 2\% level for Nyx and \CRKHACC.dtfe, while only 5-10\% for \CRKHACC.sph; all three point to the same answer, with differences $\leq 2$\% between H\_L10R10 and N\_L10R10. We note that the improvement between \CRKHACC.sph and \CRKHACC.dtfe is larger at $z=5$ since the DTFE algorithm better represents low density regions, the origin of the \lya\ signal at $z=5$, as opposed to the lower redshift signal that originates from low- to mid-density regions.

\subsection{1D power spectrum}

Spatial correlations of the \lya\ flux constitute an extremely powerful cosmological probe of matter density fluctuations at small scales (as small as 1~Mpc) and high redshifts. This increasingly used 2-point statistic has given the strongest constraints to date on neutrino masses and dark matter models~\citep{Yeche2017,irsic_first_2017,armengaud_constraining_2017,PDB2019} and has also produced competitive measurements on the IGM thermal history~\citep{walther_new_2018,Boera2019,Gaikwad2021}. 

The first category of $P_{\rm 1D}$ measurements have been performed using tens of thousands of medium-resolution SDSS quasar spectra~\citep{PalanqueDelabrouille2013,Chabanier2019a}. They measure flux fluctuations on scales $\rm 0.0015 \ km^{-1}s \leq k \leq \ 0.02 ~km^{-1}s$ with great precision, down to 1.5\% at low redshifts and large scales, thanks to the extremely favorable statistics of the quasar samples. However they miss the small-scale flux fluctuations, important to study alternative DM models or the IGM thermal history, due to lack of resolution and low SNR. By increasing the number of \lya\ quasar spectra by a factor of 4 and improving the resolution and noise, the on-going DESI survey will probe scales smaller than its predecessor, in the range $\rm 0.0015\ km^{-1}s \leq k \leq 0.04 \ km^{-1}s$, with percent level precision at low redshifts.

The second category used tens to hundreds of quasar spectra~\citep{viel_warm_2013,irsic_lyman-alpha_2017,Yeche2017,walther_new_2018,Boera2019,Gaikwad2021,Karacayli_2022} with high resolution and high S/N allowing them to probe smaller scales than the former category but without the large scales required to constrain cosmology, with scales in the complementary range $\rm 0.02~km^{-1}s \leq k \leq 0.1~km^{-1}s$. The higher resolution comes at the expense of a higher exposure time resulting in smaller data sets, and therefore a lower precision, i.e., 5-15\%.

\begin{figure*}
\centering
\includegraphics[width=0.49\textwidth]{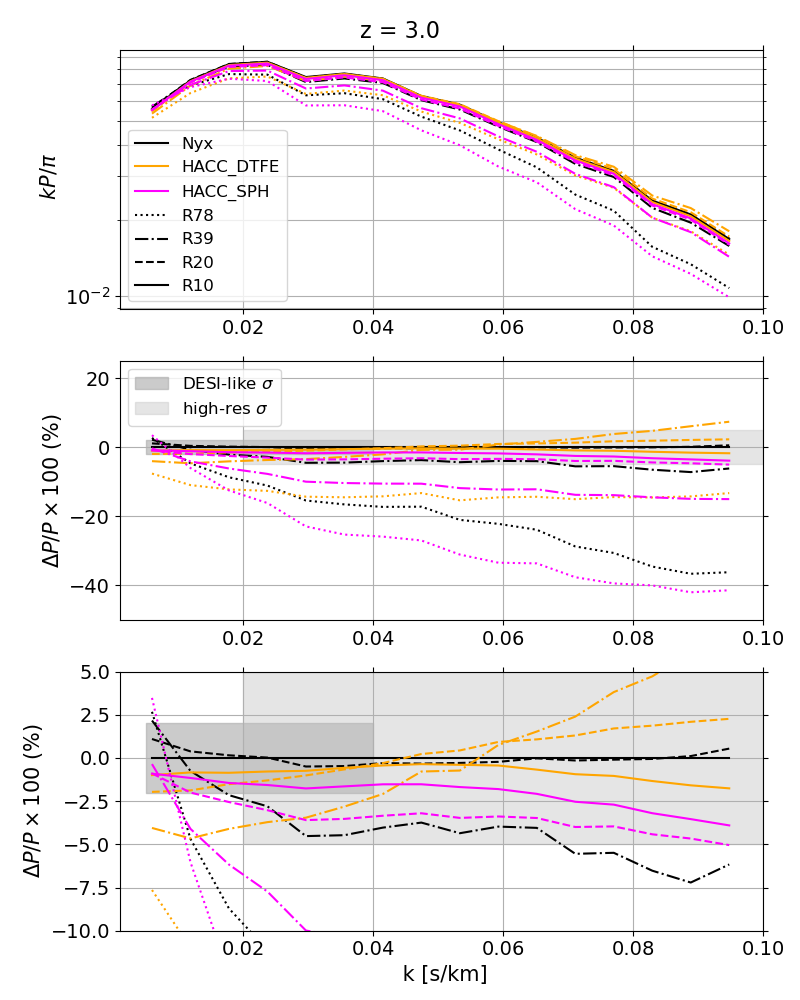}
\includegraphics[width=0.49\textwidth]{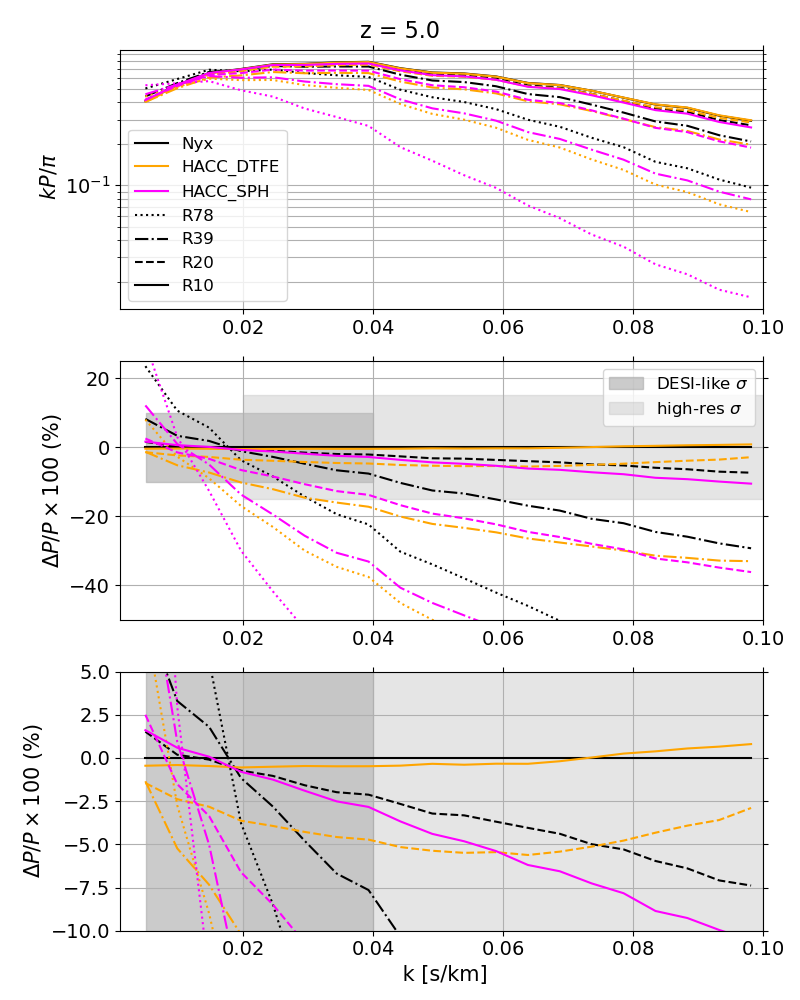}
\caption{1D Ly$\alpha$ power spectrum for Nyx (black lines) and \CRKHACC (orange and magenta lines for the DTFE and SPH density estimation, respectively) at $z=3$ (left) and $z=5$ (right) in the 10~$\rm h^{-1}Mpc$ boxes for different spatial resolutions. Middle and bottom panels show relative differences with respect to N\_L10R10, the most-resolved Nyx simulation with 10~$\rm h^{-1}kpc$ resolution. The shaded regions show $k$-averaged observational uncertainties of  DESI-like (darker gray) and high-resolution (lighter gray) surveys. Note the excellent agreement between the high-resolution Nyx and \CRKHACC results, when using DTFE with the latter.}
\label{fig:p1d}
\end{figure*}

Analogously to density fluctuations, we can define the flux perturbations as $\delta_F = F/\left< F\right >-1$, where $\left< F\right >$ is averaged over all skewers. Then, the 1D \lya\ flux power spectrum, $P_{\rm 1D}$ is obtained by Fourier transforming $\delta_F$ along each line of sight and averaging in $k_{\parallel}$. We display the dimensionless $P_{\rm 1D}$ in Fig.~\ref{fig:p1d} at $z=3$ and at $z=5$ in the 10~$\rm h^{-1}Mpc$ boxes. The shaded areas indicate observational uncertainties averaged over the wavelength range for DESI-like surveys and high-resolution surveys in darker and light gray respectively. DESI-like uncertainties cover the large-scale range $0.005 \rm{ \ km^{-1}s} \leq k \leq 0.04 \rm{ \ km^{-1}s}$ and are from the forecast of~\cite{Valluri2022}, which is based on eBOSS uncertainties~\citep{Chabanier2019a}, with a factor 4 increase in the number of quasars and improved resolution. We use the $z=4.6$ forecast for redshift $5.0$, given that 4.6 is the highest redshift bin available in the eBOSS measurement. For the high-resolution uncertainties, we use those from~\cite{Karacayli_2022} and~\citep{Boera2019} for $z=3$ and $z=5$ respectively, in the $0.02 \rm{ \ km^{-1}s} \leq k \leq 0.1 \rm{ \ km^{-1}s}$ wavelength range. In the following paragraphs, we first compare the convergence of Nyx with that of \CRKHACC.sph, then we introduce the results of \CRKHACC.dtfe. Finally, we investigate how results evolve when artificially increasing the spatial resolution using Richardson extrapolation.

\subsubsection{Comparison of Nyx and \CRKHACC.sph outputs}
\label{comp_sph}

For both codes and as a function of redshift, lower-resolution runs tend to have more large-scale power, with a larger tilt, so that they fall off at smaller scales than higher-resolution runs. These larger  discrepancies at small scales are expected because flux fluctuations are more sparsely sampled so there is less power for the low-resolution runs. Similarly for the previous flux statistics, the increase in spatial resolution requirement with redshift to model the \lya\ $P_{\rm 1D}$ is because of the larger filtering scale at high redshift, the signal originating from lower-density regions, and of weaker thermal broadening.

At $z=3$, the convergence is again faster for Nyx, indeed N\_L10R20 and N\_L10R10 are almost exactly the same at all scales, even as small as 0.1~km$^{-1}$s proving that N\_L10R20 is already fully converged, in agreement with previous findings~\citep{Tytler2009,Lukic2015}. \CRKHACC.sph is slower to converge as there are still differences between H\_L10R20 and H\_L10R10, at the $\sim$ 2\% level. 
When comparing the most resolved power spectra, the agreement between Nyx and \CRKHACC.sph is at the limit of observational uncertainties. For large scales, $k \leq 0.04$~km$^{-1}$s, probed by medium resolution surveys, Nyx and \CRKHACC runs tend to the same answer at the 1.5\% level, and disagree at only $< 5$\% at small scales, i.e. $0.04 \leq k/{\rm km^{-1}s} \leq 0.1$, probed by high-resolution surveys.

At $z=5$, the convergence is significantly slower for both codes, with better convergence for Nyx at all scales. Agreement between N\_L10R20 and N\_L10R10 is better than 2.5\% on large scales and better than 7.5\% on small scales. In contrast, agreement between H\_L10R20 and H\_L10R10 is only at 10\% on large scales and 30\% on small scales. Finally, the agreement between Nyx and \CRKHACC.sph at the best spatial resolution is better than current observational uncertainties, but still quite large, about 2.5\% on large scales and 10\% on small scales.

Finally, as discussed in the previous sections, \CRKHACC.sph requires more particles to be converged, i.e., H\_L10R10 is converged but H\_L10R20 is not, as opposed to Nyx for which 20~$\rm h^{-1}kpc$ resolution is enough to be converged with respect to current observational uncertainties. Also, at equivalent average resolution, we find that \CRKHACC.sph has a systematically lower \lya\ $P_{\rm 1D}$ compared to Nyx. This is in accordance with the previous results shown for the individual skewers (Fig.~\ref{fig:skewers}) or the effective resolution as a function of density (Fig.~\ref{fig:reso}) from which it is clear that \CRKHACC.sph computes systematically higher densities in low-density regions and is thus more opaque to \lya\ transmission. Nevertheless, we find that the differences between \CRKHACC.sph and Nyx at equivalent average resolution decrease with increasing resolution -- they are at the limit of the observational uncertainties when comparing the best-resolution runs. The requirement of 10~$\rm h^{-1}kpc$ average resolution for \CRKHACC.sph at $z=3$ looks reasonable from Fig.~\ref{fig:reso} given that the effective resolution is about 20~$\rm h^{-1}kpc$ on the range of density producing the \lya\ signal, as shown with the black dashed line (\CRKHACC.sph resolution) lying below the magenta solid line (20~$\rm h^{-1}kpc$ constant resolution of Nyx). For $z=5$ though, precision modeling would require an average resolution of 5~$\rm h^{-1}kpc$ at least.

If the difference between the two codes at the best resolution is at the limit of observational uncertainties, there are still  1.5\% scale-independent and scale-dependent biases at large and small scales respectively. We perform a series of tests to investigate the sources driving these discrepancies.

\begin{itemize}
\item We lower the numbers of neighbors to  $n_{\rm nb} = 113$, instead of $n_{\rm nb} = 268$ for the low-resolution run H\_L10R39. This improves the results by 2\% and 10\% at most at small scales at $z=3$ and $z=5$ (see App.~\ref{sec:nb}). Therefore, we do not expect this parameter to significantly alter the results at higher resolution. 
\item We take the Nyx velocity instead of the \CRKHACC.sph velocity to compute the flux power spectra and the results remain unchanged. 
\item If the $\gamma$ thermal parameters are very close for  H\_L10R10 and  N\_L10R10, the temperature at mean density, $T_0$, shows a few percent difference (see Fig.~\ref{fig:t0gamma}). We compute the flux power spectra for H\_L10R10 by rescaling the temperature and density so that the new $T_0$ and $\gamma$ match that of N\_L10R10 but it shows very little change (0.5\% at the smallest scales at $z=3$). 
\item We remove the effect of high density regions, since we see in Figs.~\ref{fig:skewers} and \ref{fig:slices} that these regions are quite different, notably because of the activation of collisionless star particles in \CRKHACC.sph and the difference in the velocity field. We remove these regions by thresholding to densities less than 50 times the mean density. If the density of a cell exceeds $50 \bar{\rho_b}$, we set $\rho_b = 50 \bar{\rho_b}$ and set the temperature of the gas using the fluctuating Gunn-Peterson approximation-like equation of state following Eqn.~\ref{eq:tempdensity}, with $T_0$ and $\gamma$ matching the simulation fit. We then recompute the optical depth and \lya\ flux the regular way, and the resulting flux power spectra show no significant deviation. 
\item From Figs.~\ref{fig:skewers} and \ref{fig:reso}, showing skewers and the resolution as function of density in the simulations, we observe biased sampling in void regions for \CRKHACC.sph. To assess the impact of this discrepancy on the flux power spectrum, we use the Nyx density instead of the \CRKHACC.sph density, when the latter is below a certain threshold, $\rho_{\rm thresh}$, and we recompute the optical depth the regular way. We gradually increase $\rho_{\rm thresh}$ from $-0.75$ to $0.5$ in logarithmic units of $\bar{\rho}_{b}$. We find that values of $\log(\rho_{\rm thresh}/\bar{\rho}_{b}) = -0.5$ and $0.5$ for $z=5$ and $z=3$ respectively, are enough to make the \CRKHACC.sph and Nyx results agree at better than 1\% at all scales, confirming that the low density regions are responsible for these remaining discrepancies. These results and further discussion are available in App.~\ref{sec:densitythresh}.
\end{itemize}

\subsubsection{Comparison with \CRKHACC.dtfe}

\begin{figure*}
\centering
\includegraphics[width=0.49\textwidth]{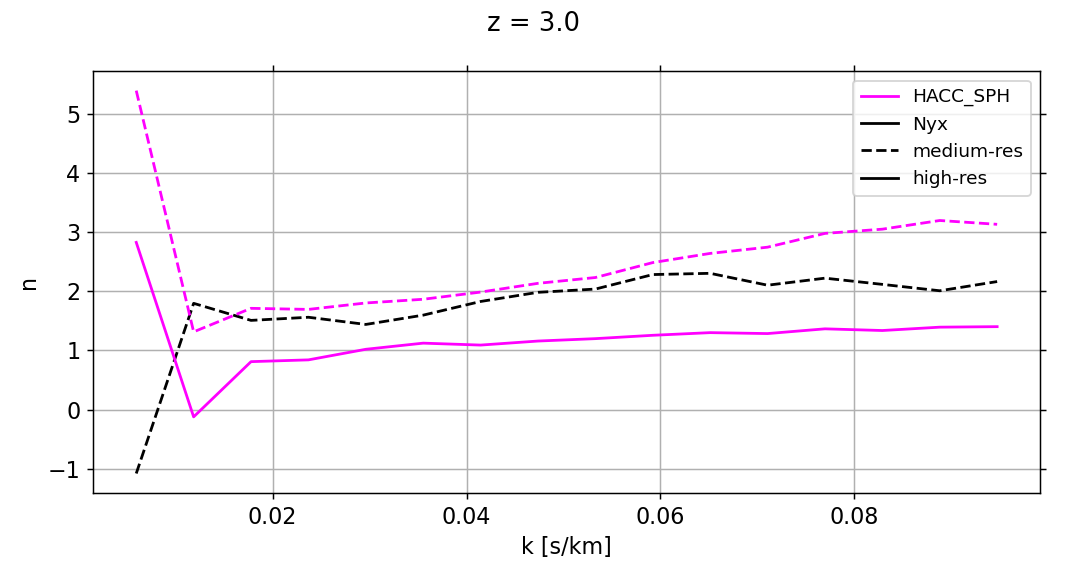}
\includegraphics[width=0.49\textwidth]{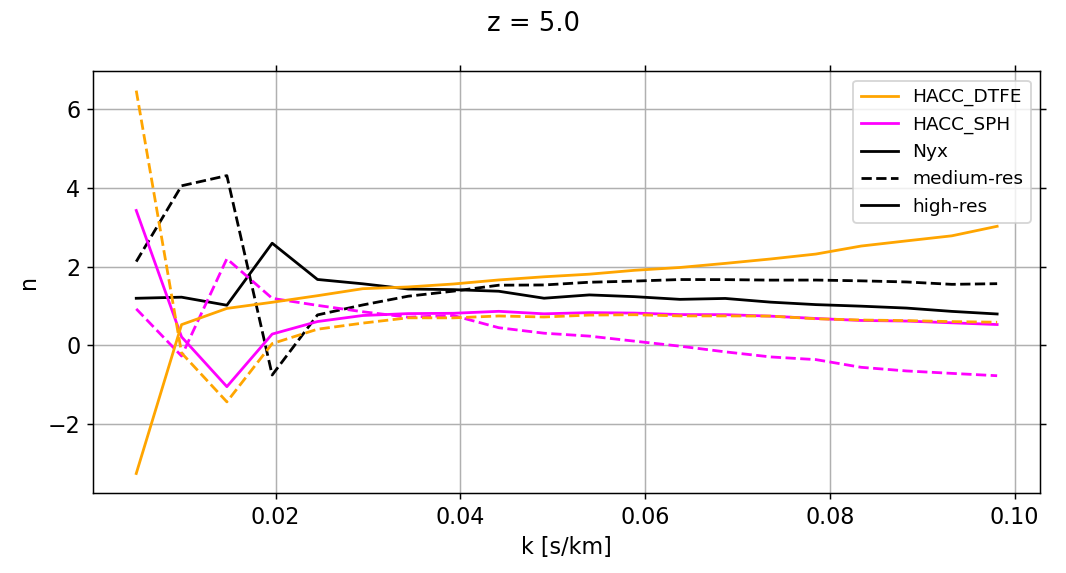}
\caption{Order of accuracy $n$ at $z=3$ (left) and $z=5$ (right) for Nyx (black) and \CRKHACC runs using the SPH (magenta) and DTFE (orange) density field estimators. The power-law index was estimated using the sets of medium resolution (L10R78, L10R39 and L10R20) and high resolution (L10R39, L10R20 and L10R10) simulations (dashed and solid  lines, respectively). For more details, see text (Section~\ref{sec:richardson}).
}
\label{fig:richardson}
\end{figure*}

The realization that the low-density regions are responsible for the main discrepancies between Nyx and \CRKHACC.sph is the driving motivation for using DTFE in order to improve the sampling of low-density regions. The corresponding \lya\ $P_{\rm 1D}$ results for \CRKHACC.dtfe outputs are shown in Fig.~\ref{fig:p1d} in orange.

We first observe that, as opposed to Nyx and \CRKHACC.sph, the convergence of \CRKHACC.dtfe is non-monotonic at $z=3$ as the different power spectra oscillate around the final most converged power spectrum, reflecting the DTFE sensitivity (as a local method) to particle sampling noise. The \CRKHACC.dtfe power spectra have some times more power at small scales than the most resolved Nyx power spectrum. Also, the convergence is  faster to reach for \CRKHACC.dtfe than for \CRKHACC.sph, especially at $z=5$.

 More importantly, the most converged power spectra of Nyx and \CRKHACC.dtfe agree on the whole range of scales at the percent level at $z=3$ and at the sub-percent level at $z=5$. Power spectra of L10R20 of Nyx and \CRKHACC.dtfe already agree at 2\% at $z=5$, whereas they are not fully converged with respect to spatial resolution. Unlike \CRKHACC.sph, an average resolution of 20~$\rm h^{-1}kpc$ appears sufficient to consider the \lya\ power spectra as being converged at the percent level at $z=3$ and at a few percent at $z=5$, considerably decreasing the computational time needed to reach the required precision given by observational surveys. We reach a better agreement between Nyx and \CRKHACC.dtfe at $z=5$ since the \lya\ signal comes from lower density regions at $z=5$ than at $z=3$, due to the improvement from using the DTFE algorithm. Note that this did not involve any fine-tuning of parameters, neither for the simulation run nor for the analyses, to get the results in such remarkable agreement. This is to our knowledge the first time that \lya\ statistics results from grid-based and SPH hydrodynamical simulations have been brought to such an extreme level of agreement, thanks to the better sampling of low density regions with the DTFE algorithm, which strongly increases our confidence that both codes are converging to the correct answer.

\subsubsection{Richardson extrapolation}
\label{sec:richardson}

In this section we study the agreement between the codes and their convergence properties when extrapolated to the ``continuum'' solution using Richardson extrapolation. This technique has been used to study the convergence properties of cosmological simulations where a priori predictions are difficult to make or not available (e.g., \citealt{Lidz2005, 2010ApJ...715..104H}, \citealt{Lukic2015}), but for completeness we repeat the main reasoning here.
A numerical method that is $n$-th order accurate in space produces a numerical approximation $P(h)$, where $h$ is the discretization element, such that
\begin{equation}
P(h) = P +  Ah^n + O(h^{n+1}),
\end{equation}
with $P$ being the exact value, $Ah^n$ the leading error, and the last term the higher-order error.  In general, the order of accuracy, especially for multi-dimensional, multi-physics codes like those considered here, is best derived from the numerical solutions themselves, even given {\em a priori} expectations for the convergence.  Using three numerical solutions for $P$ with three spatial resolutions $h$, $rh$ and $r^2h$, where $r$ is the constant refinement ratio, and {\em assuming that convergence is uniform}, the order of accuracy is:
\begin{equation}
n = \frac{\ln \left(   \frac{P(r^2h)-P(rh)}{P(rh)-P(h)} \right)  }{\ln(r)} \, .
\end{equation}
Knowing $n$, Richardson extrapolation allows to extrapolate $P$ to the $n$-th order of accuracy ``continuum'' solution, $h \to 0$, using two numerical results for $P$:
\begin{equation}
P_R =   \frac{r^n P\left( h\right) - P\left( r h\right) }{r^n-1} \, .
\end{equation}
\label{eq:richardson}
This ``extrapolation to the limit" is usually checked at three resolution values to verify the value of $n$, before the final extrapolation is implemented.

To quantitatively assess the convergence of $P_{\rm 1D}$, we computed the order of accuracy, $n$, along with the differences between the 1D Ly$\alpha$ power spectra derived from the simulation, $P_{\rm N}$ and $P_{\rm H}$ for Nyx and \CRKHACC respectively, and the power spectra derived from the Richardson extrapolation, $P_R$, at $z=3$ and $z=5$, as shown in Fig.~\ref{fig:richardson}.  Note that we do not include results for \CRKHACC with DTFE outputs at $z=3$ since Richardson extrapolation requires monotonic convergence which is not the case with DTFE as shown in Fig.~\ref{fig:p1d}.

\begin{figure*}
\centering
\includegraphics[width=0.49\textwidth]{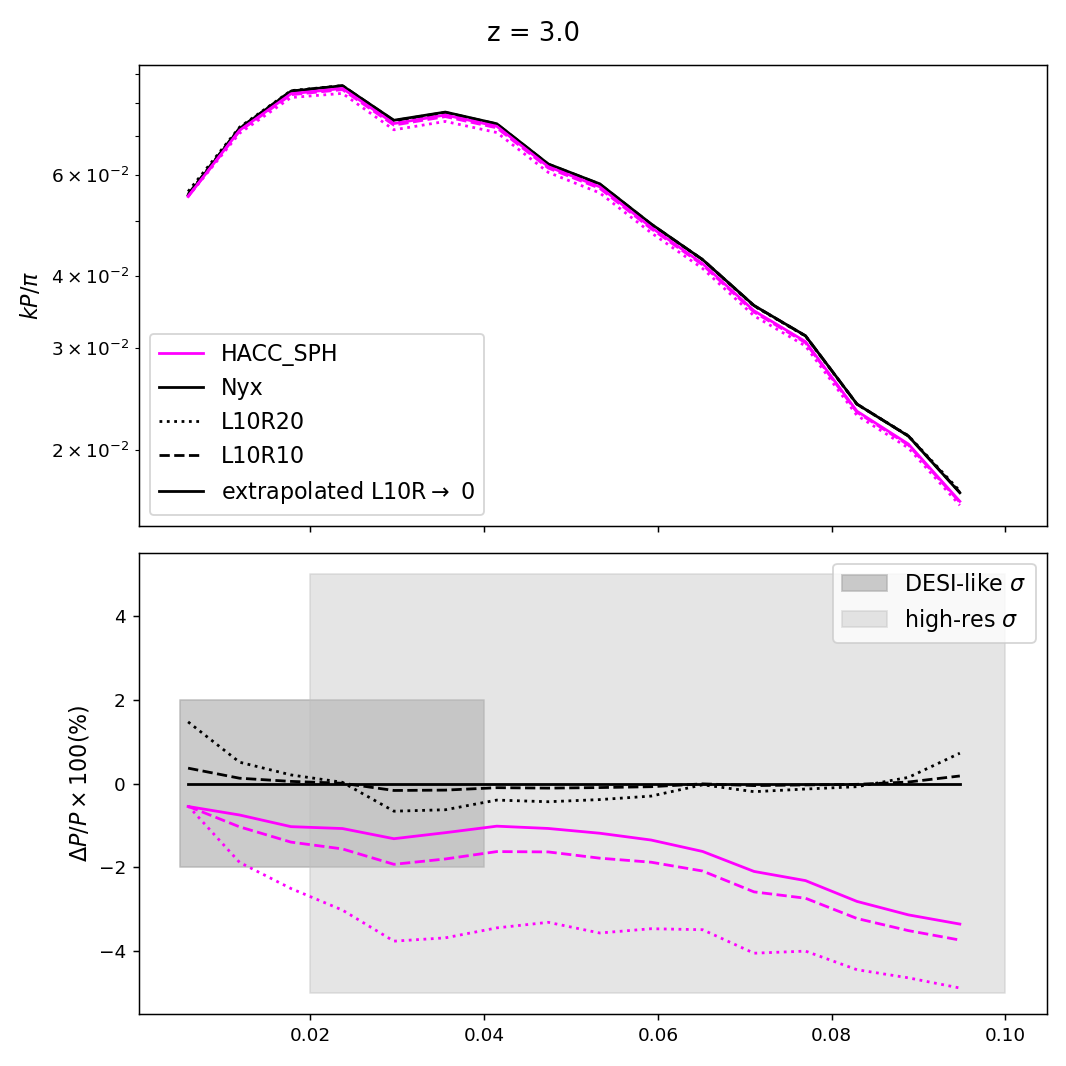}
\includegraphics[width=0.49\textwidth]{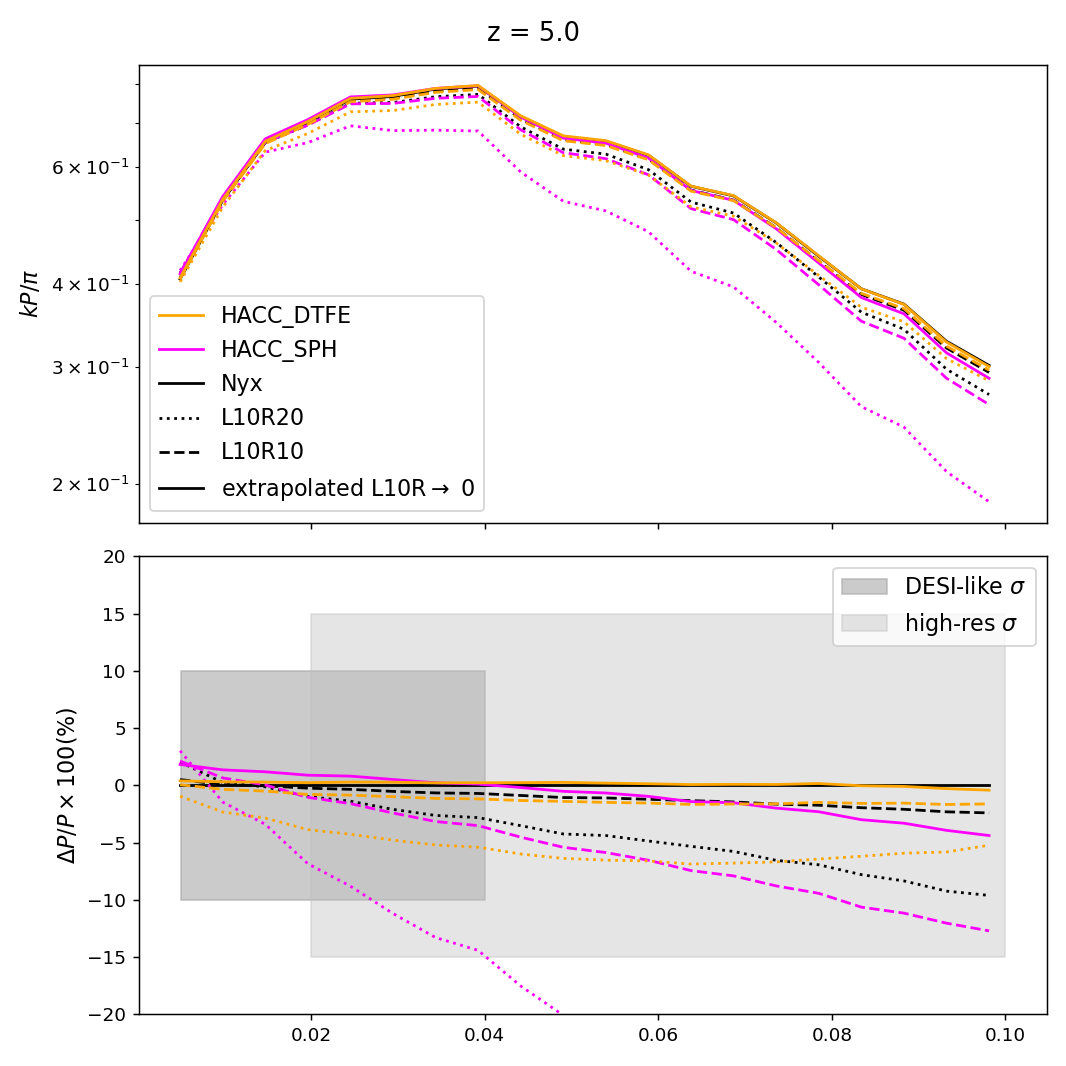}
\caption{1D Ly$\alpha$ power spectrum for Nyx and \CRKHACC.sph at $z=3$ (left) and $z=5$ (right) in the 10 $\rm h^{-1}Mpc$ boxes for different spatial resolution. Middle and bottom panels show relative differences with respect to N\_L10R0, the Richardson extrapolated power spectra with resolution $R \rightarrow 0$.}
\label{fig:p1d2048}
\end{figure*}

We use the medium-resolution set, composed of the L10R78, L10R39 and L10R20 runs, and the high-resolution set, composed of the L10R39, L10R20 and L10R10 runs, to compute $n_{\rm medium-res}$ and $n_{\rm high-res}$ and assess how it evolves with increasing spatial resolution. At $z=3$, we do not display $n_{\rm high-res}$ for Nyx because $P_{\rm N\_L10R10}$ and $P_{\rm N\_L10R20}$ are almost exactly matching (see Fig.~\ref{fig:p1d}) and thus the calculation for $n$ becomes dominated by sub-percent numerical noise differences between the two runs.
For Nyx, at both $z=3$ and $z=5$, $n$ is in remarkable agreement with the theoretical value $n=2$. For \CRKHACC, at $z=3$ $n$ is also nicely converging to $n=2$, with a slightly lower value than Nyx. At $z=5$, for \CRKHACC.sph, $n$ is negative when using the medium resolution set and tends toward 1 when using the high-resolution set, which is not surprising because the power spectrum is quite far from being properly convergent at this redshift. For \CRKHACC.dtfe, $n_{\rm medium-res}$ almost exactly matches $n_{\rm high-res}$, and $n_{\rm high-res}$ is in very good agreement with the $n=2$ theoretical value, except at very small scales where it diverges toward high values.

Finally, to further assess the convergence and agreement with respect to the spatial resolution for both Nyx and \CRKHACC.sph, we use Richardson extrapolation to extrapolate power spectra to the limiting case of an infinitely small resolution element.
We show the results at $z=3$ and $z=5$ in Fig.~\ref{fig:p1d2048}. At $z=3$, as expected for Nyx, extrapolation does not improve the result since $P_{\rm N\_L10R20}$ and $P_{\rm N\_L10R10}$ are already almost identical. For \CRKHACC.sph, results are hardly improved with differences below 1\% at all scales between $P_{\rm H\_L10R10}$ and $P_{R,\rm H\_L10R0}$  while $P_{\rm H\_L10R10}$ and  $P_{\rm H\_L10R20}$ are a few percent different confirming that we need an average resolution of 10~$\rm h^{-1}kpc$ for \CRKHACC.sph to reach the 1\% convergence at $z=3$ at scales as small as 0.1~km$^{-1}$s. It is important to note however that the $P_{1D}$ of the extrapolated Nyx and \CRKHACC.sph runs, N\_L10R0 and H\_L10R0, still display a 1\% systematic offset at large scales and scale-dependent bias at small scales up to 2\%. But these are at the limit of current observational uncertainties, and also appear to be degenerate with the mean flux on large scales, hence unlikely to impact cosmological parameter inference.  At $z=5$, Richardson extrapolation improves the results by 2\% at small scales, compared to 10\% discrepancies between $P_{\rm N\_L10R20}$ and $P_{\rm N\_L10R10}$. This shows that while there is room for improvement, the errors are still significantly below the observational uncertainties, which are about 10-15\% at high redshifts. This demonstrates that 10~$\rm h^{-1}kpc$ resolution is enough for Nyx, or presumably any grid-based code, at $z=5$. $P_{R,\rm H\_L10R0}$ displays up to 8\% more power compared to $P_{\rm H\_L10R10}$, and it is even closer to the Nyx results, with agreement at the 2\% level even at small scales.

The results confirm that Richardson extrapolation works well to estimate simulated power spectra to a higher resolution than the native resolution of the simulations. In practice, we could use 80~$\rm h^{-1}kpc$ and 40~$\rm h^{-1}kpc$ resolution simulations to extrapolate a converged power spectra for Nyx instead of running a 20~$\rm h^{-1}kpc$ simulation at redshifts $z \sim 3$. For HACC, if using the DTFE density field estimator, 80~$\rm h^{-1}kpc$ and 40~$\rm h^{-1}kpc$ resolution simulations are also sufficient to extrapolate a converged power spectrum, whereas if using the SPH density estimator, 40~$\rm h^{-1}kpc$ and 20~$\rm h^{-1}kpc$ averaged resolution simulations are required to extrapolate to a converged result.

\begin{figure*}
\centering
\includegraphics[width=1.00\textwidth]{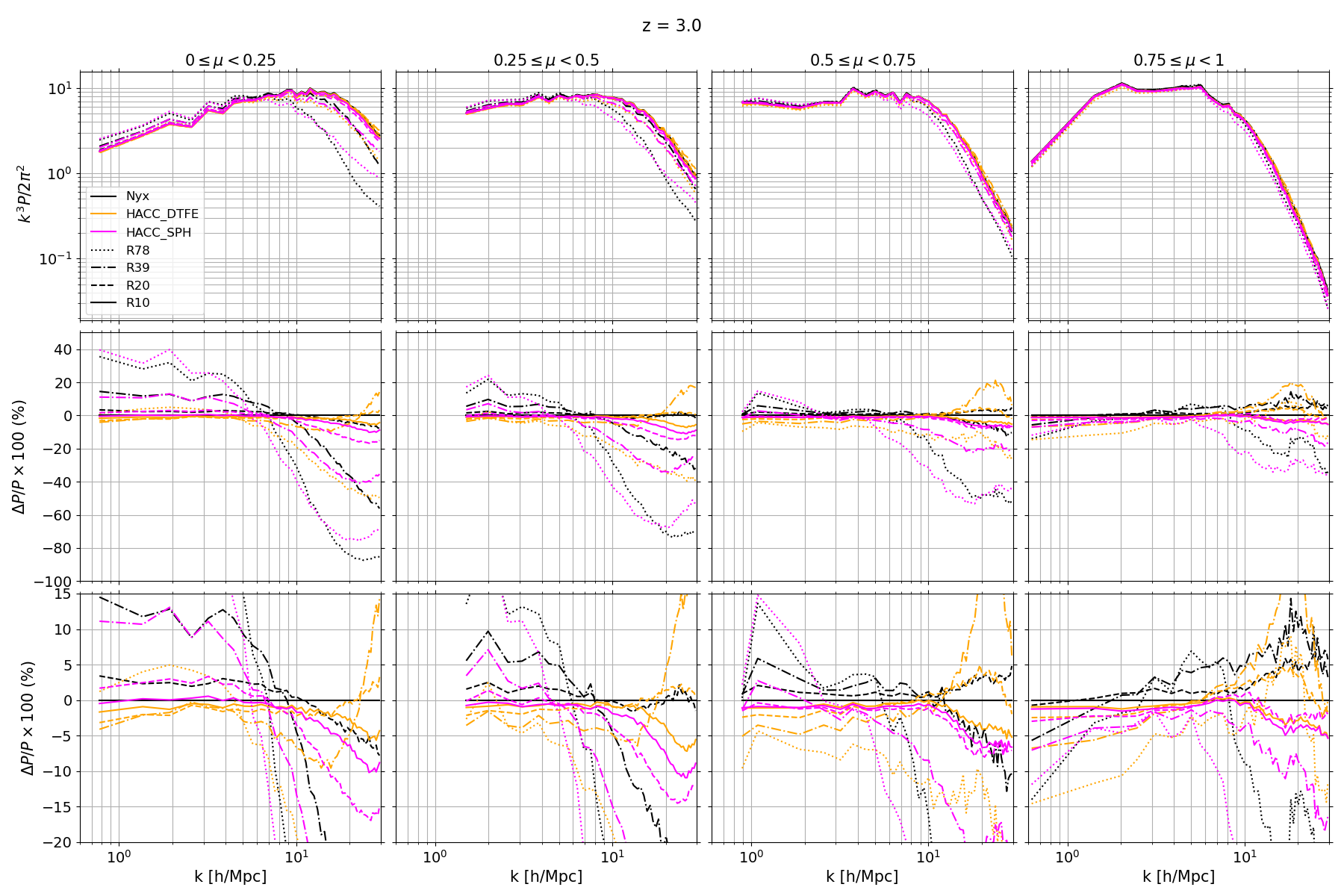}
\caption{3D Ly$\alpha$ power spectra at $z=3$ for Nyx (black line) and \CRKHACC (orange and magenta lines for the DTFE and SPH density estimation, respectively) in the 10~$\rm h^{-1}Mpc$ boxes for different spatial resolutions. Middle and bottom panels show relative differences with respect to N\_L10R10, the most-resolved Nyx simulation with 10~$\rm h^{-1}kpc$ resolution. The bottom panel shows a more restricted range on the $y$-axis to make it easier to distinguish the different results.}
\label{fig:p3d}
\end{figure*}

\subsection{3D power spectrum}
\label{sec:3dpspec}

In contrast to the 1D Lyman-${\alpha}$ power spectra that yield \lya\ correlations only along the LOS, the 3D  Lyman-${\alpha}$ power spectra, $P_{\rm 3D}$,  gives \lya\ correlations in all directions. The two are related by
\begin{equation}
    P_{\rm 1D}(k_{\parallel}) = \int ^{\infty}_0 P_{\rm 3D}(k_\perp,k_\parallel) \frac{dk_\perp k_\perp}{2 \pi}.
\end{equation}
The $P_{\rm 3D}$ has not been measured with observational data yet because of lack of surveys with high enough LOS density. Since the quasar density for DESI will allow such measurements, algorithms have already been developed to estimate $P_{\rm 3D}$ with real data~\citep{FontRibera2018}.  As such, it is timely to prepare the theoretical modeling of these statistics. Here, we only show results at $z=3$ given that the first measurement will be performed at low redshifts, where the quasar density is the highest. We show the equivalent results at $z=5$ in App.~\ref{sec:p3dhighz}.

To visualize the anisotropy of the power spectrum field we follow the convention of using $P_{\rm 3D}(k,\mu)$, where $\mu$ is the cosine of the angle between the mode and the LOS, i.e., $\mu = k_\parallel/k$. We average the power spectrum in 4 $\mu$ bins. The resulting 3D flux power spectra are shown in Fig.~\ref{fig:p3d}. From left to right, the power $\mu$ wedges shown are increasingly parallel to the LOS. That is, the panel on the left can be thought of as transverse power and the panel on the right as LOS power, i.e., close to $P_{\rm 1D}$.

The observations on resolution and redshift evolution trends of the $P_{\rm 1D}$ section still hold and can be generalized for every $\mu$ bin. For Nyx and \CRKHACC.sph, lower-resolution runs have more power at large scales and less power at small scales compared to high-resolution runs because flux fluctuations are more sparsely sampled (the discrepancies are larger at high redshifts, see Fig.~\ref{fig:p3d_highz}). In contrast, the lower-resolution power spectra from \CRKHACC.dtfe display a different sense of convergence as a function of $k$, with less power at large scales and higher power at small scales. We also observe that the LOS power (right panel) tends to agree much better across resolutions and codes. This is due to the thermal broadening along the LOS, which acts to smooth out fluctuations in this direction. The perpendicular perturbations are not filtered in this way, and show much more small-scale power, which enhances discrimination between simulation results run at  different resolutions.

The agreement between the most resolved results is remarkably good for all $\mu$ bins on the range of scales probed by DESI-like surveys, i.e. up to 10 $\rm h^{-1}Mpc$. Surprisingly, the systematic $\sim$ 1\% offset observed on the largest scales for the $P_{\rm 1D}$ with \CRKHACC.sph (see Fig.~\ref{fig:p1d}) disappears as we move to a direction perpendicular to the LOS.

\section{Box size effects}
\label{sec:boxsize}
The discussion so far has been limited to the study of flux statistics from the small-box simulations. Although the L10R20 and L10R10 simulations appear to be converged with respect to resolution, they are definitely not converged in box size. By redshift $z=2.0$, the box-scale mode is significantly nonlinear for a 10~h$^{-1}$Mpc box size, which accounts for a certain amount of missing power in the simulations. Additionally, the missing large-scale modes also significantly affect the properties of bulk flows in the simulations. This is generally the least converged quantity in small-box simulations, and results in a lower temperature distribution (affecting the broadening of the lines) and smaller peculiar velocities (affecting the redshift-space distortions). In order to compare more representative predictions from both codes, we also consider statistics from larger box simulations, in a box of 40~$\rm h^{-1}Mpc$ of side, with resolution up to 20~$\rm h^{-1}kpc$, since previous studies show that this box-size is enough to avoid finite-size effects~\citep{McDonald2003, Lidz2005, Tytler2009, Lukic2015}.

We find results similar to~\cite{Tytler2009} and \cite{Lukic2015} when comparing results between the L10R20 and L40R20 runs for Nyx and \CRKHACC (see App.~\ref{sec:largebox}). In particular, the mean flux is larger for the bigger box, with more enhanced discrepancies at high redshift, at the 3\% level at most. For the 1D \lya\ power spectrum, in addition to not probing scales as large as the L40 runs, the L10 runs display up to 10\% discrepancies with the L40 \lya\ power spectra at both redshifts.

More importantly, in the context of this paper, when comparing differences between \CRKHACC and Nyx results for simulations in the 40~$\rm h^{-1}Mpc$ boxes with varying spatial resolution, we reach similar conclusions as in the case of the 10~$\rm h^{-1}Mpc$ boxes (Section~\ref{sec:flux}). Using the SPH density field for \CRKHACC, the Nyx results converge faster for all the flux statistics and the differences between the two codes when comparing the most converged results are the same. In particular, the offsets between the two power spectra at $z=3.0$ and $z=5.0$ are identical; thus the level of agreement between Nyx and \CRKHACC in terms of \lya\ statistics does not depend on the box size.


\section{Conclusions}
\label{sec:conclusion}
In this paper, we compare \lya\ forest statistics in numerical simulations using two very different hydrodynamical solvers, the Lagrangian SPH \CRKHACC code~\citep{Habib2016,Frontiere2017} and the Eulerian grid-based Nyx code~\citep{Almgren2013, Sexton2021}. The primary aim is to establish a level of confidence in theoretical predictions extracted from hydrodynamical simulations to interpret the data from future or ongoing \lya~surveys, such as DESI~\citep{DESI2016} or WEAVE-QSO~\citep{Pieri2016}.

We used a set of 12 main simulations, comprising of 6 \CRKHACC and 6 Nyx runs, and 7 supporting \CRKHACC simulations, in box sizes ranging from 10 to 40~$\rm h^{-1}Mpc$ with number of cells/particles ranging from $128^3$ to $2048^3$, translating into average resolutions ranging from 78 to 10 ~$\rm h^{-1}kpc$. Comparing results between the two codes was performed on the same Cartesian grid, which required a choice of density estimation for the \CRKHACC results. We used two approaches --  the standard SPH method where density is averaged over all neighbors in a smoothing sphere (\CRKHACC.sph results) and the Delaunay Tesselation Field Estimator (\CRKHACC.dtfe results) for which the volume is covered by contiguous and non-overlapping tetrahedra.

Morphological comparisons of line-of-sight skewers and 2D slices show good agreement between the two codes. We find two main discrepancies between the \CRKHACC.sph and Nyx outputs, with the former overestimating the density in low-density regions and missing small-scale fluctuations because of the inherent smoothing nature of the SPH density estimator, as seen in Fig.~\ref{fig:skewers}. The use of the DTFE estimator significantly improves the comparisons, as we discuss later below.

We compute the \lya\ flux field for all outputs and compare the 0-, 1-, and 2-point flux statistics, i.e., the mean flux, the flux probability distribution function and the flux power spectrum, the most commonly used statistics to extract astrophysical and cosmological information from observations. For all statistics, both codes have  similar convergence trends with higher resolution requirement and larger discrepancies between the codes at high redshifts and/or at small scales, the former being due to the fact that at high redshifts the Jeans length is larger, thermal broadening weaker, and the \lya\ signal originates from lower density regions. 

We find that the Nyx results converge significantly faster than \CRKHACC.sph, requiring less cells than the latter to predict converged \lya\ flux statistics. This is partly due to the intrinsically density-adaptive nature of Lagrangian codes, since particles are attracted to high-density regions at the expense of low-density regions making it harder to control resolution in the low-density regime, without increasing the particle sampling density. This behavior is shown in Fig.~\ref{fig:reso}, which presents the effective resolution as a function of density in simulations varying the number of cells/particles. Agreements between the converged predictions of Nyx and \CRKHACC.sph are summarized below:
\begin{itemize}
    \item Mean flux: better than 1\% for $2 \leq z \leq 3$ and better than 2.5\% for $z\leq 5$
    \item Flux PDF: better than 5\% at $z=3$ and better than 30\%  for $F \leq 0.8$ at $z=5$ since low-flux values are not produced in the \CRKHACC.sph output
    \item Flux 1D power spectrum: better than 2.5\% and 5\% at large and small scales respectively at $z=3$, which is at the limit of observational uncertainties, and better than 5\% and 10\% at large and small scales respectively at $z=5$
\end{itemize}
To understand the sources of the remaining discrepancies between Nyx and \CRKHACC.sph outputs, we test how the flux power spectrum is modified when varying the number of neighbors, $n_{\rm nb}$, taking the Nyx velocity instead of the \CRKHACC velocity to compute the \CRKHACC flux field, rescaling $T_0$ and $\gamma$ to match between the two codes, removing high-density regions and using the Nyx density field instead of the \CRKHACC density field in low-density regions to compute the \CRKHACC flux field. We find that differences in the low-density regions between Nyx and \CRKHACC outputs are responsible for the systematic discrepancies between the codes while tests for other sources for possible differences showed negligible impact.

In order to improve the sampling of the lower-density regions in \CRKHACC outputs, we use the DTFE algorithm of~\citet{Rangel2016}, yielding density fields in much better agreement with Nyx, with more small-scale fluctuations and deeper voids (see Fig.~\ref{fig:skewers}). The convergence of \lya\ flux statistics of \CRKHACC.dtfe outputs is as fast as those from Nyx; comparisons between converged predictions of Nyx and \CRKHACC.dtfe show agreement at:
\begin{itemize}
\item Mean flux: better than 1\% at all redshifts
\item Flux PDF: better than 5\% at $z=3$ and better than  than 30\%  for $F \leq 0.8$ at $z=5$ (i.e. with a upper bound for the flux 30\% higher than for \CRKHACC.sph) for the flux PDF
\item Flux 1D power spectrum: better than 1.5\% at all scales at $z=3$ and better than 1\% at all scales at $z=5$ for the flux power spectrum.
\end{itemize}

We stress that this is the first time that two completely different hydrodynamical methods, using Eulerian and Lagrangian approaches, agree on \lya\ predictions (or any cosmological observable which requires hydrodynamical modeling in the nonlinear regime!) at the percent level.  Moreover, this was achieved without any fine tuning of code parameters, raising our confidence that both methods are converging towards the correct answer.
Observational measurements of the \lya\ forest are now also reaching percent level precision, thanks to the strong increase in the number of observed quasars. Those measurements will enable much better constraints of cosmological parameters, including neutrino mass and potential non-CDM models, but only if we are able to produce theoretical models at a similar or better level of accuracy.  The comparison of computational methods we present here is, therefore, particularly timely.

In the interest of completeness, we note that at the few percent level \lya\ forest observables are also sensitive to secondary effects related to galaxy formation and evolution, such as AGN and supernova feedback~\citep{Viel2013,Chabanier2020,MonteroCamacho2021,Burkhart2022}. Incorporating these effects into  hydrodynamical simulations requires implementation of (relatively ad hoc) subgrid models, which are commonly calibrated on astrophysical observations and more strongly differ between different implementations. Such physical models were not a part of the analysis presented here. Nevertheless, we wish to emphasize that the simulations we produced, incorporating atomic cooling and UV background heating, are the simulations commonly used by the broader \lya\ community to constrain cosmological and astrophysical parameters with \lya\ observables.  We foresee that this situation will be stable in the near future, and do not expect simulation campaigns to simultaneously vary cosmological and galaxy formation parameters constraining both from the \lya\ forest.  Instead, the impact of these effects (or uncertainties associated with them) can be parameterized by running an ensemble of simulations and varying subgrid parameters (as in~\citealt{Chabanier2020} for the impact of AGN feedback) and including those as corrections in either post-processing of simulation forward models or in the likelihood calculation itself.

\section*{Acknowledgements}
We thank Eric Armengaud for the forecast of DESI osbervational uncertainties.
This research was supported by the Exascale Computing Project (17-SC-20-SC), a collaborative effort of the U.S. Department of Energy Office of Science and the National Nuclear Security Administration.  Lawrence Berkeley Laboratory's work was supported under the U.S. Department of Energy contract DE-AC02-05CH11231. Argonne National Laboratory's work was supported under the U.S. Department of Energy contract DE-AC02-06CH11357. This work used resources of the Oak Ridge Leadership Computing Facility, which is a DOE Office of Science User Facility supported under Contract DE-AC05-00OR22725. Additionally, this study utilized resources of the Argonne Leadership Computing Facility, which is a DOE Office of Science User Facility supported under Contract DE-AC02-06CH11357. This research also used resources of the National Energy Research Scientific Computing Center (NERSC), a U.S. Department of Energy Office of Science User Facility located at Lawrence Berkeley National Laboratory, operated under Contract No. DE-AC02-05CH11231 using NERSC award m3921 in 2021/22. 
\section*{Data Availability}

The data used for testing the simulation results in this paper will be made available on request.



\bibliographystyle{mnras}
\bibliography{main} 




\appendix

\section{Additional Examples and Code Tests}
\label{appendix}
\subsection{Density fields in low-resolution simulations}
\label{sec:lowres}

\begin{figure}
\centering
\includegraphics[width=1.00\columnwidth]{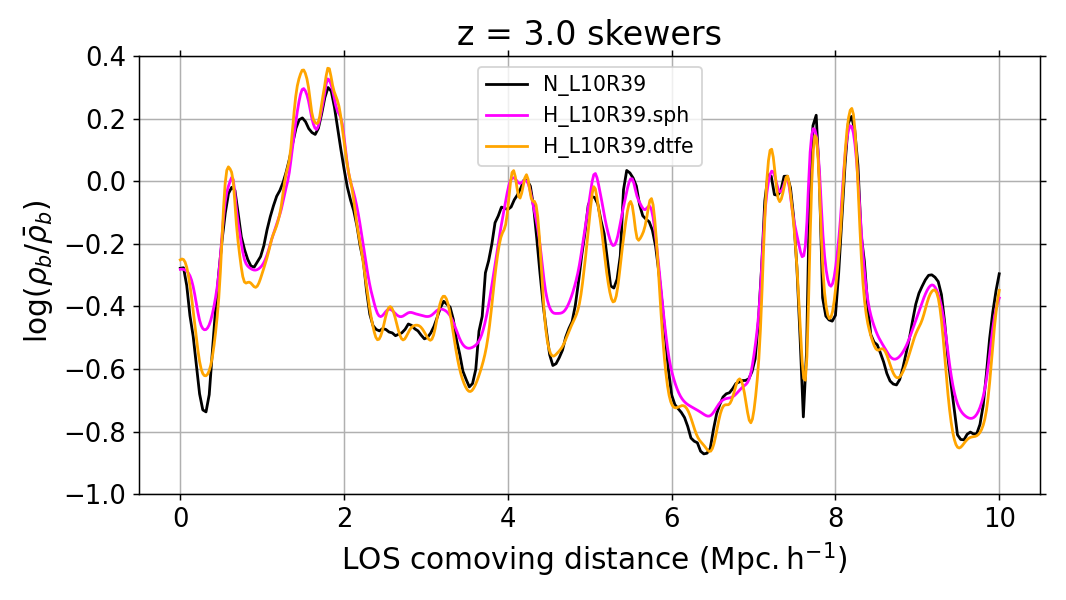}
\includegraphics[width=1.00\columnwidth]{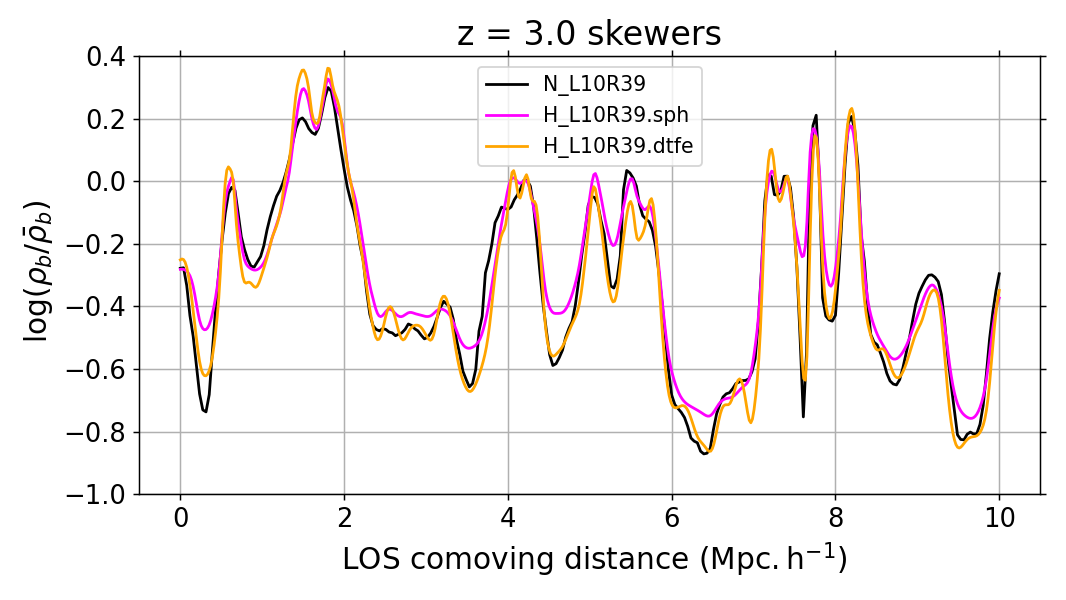}
\includegraphics[width=1.00\columnwidth]{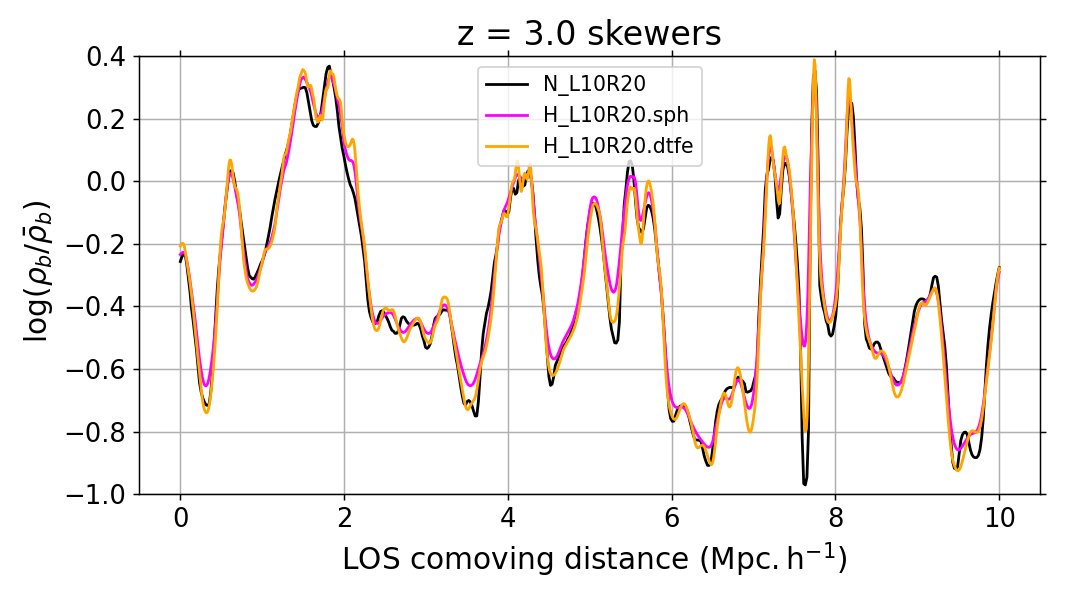}
\caption{Density field skewers from the snapshots in the 10~$\rm h^{-1}Mpc$ box with $128^3$ (top), $256^3$ (middle) and $512^3$ (bottom) cells/particles at $z=3$ for Nyx (black) and  \CRKHACC using the SPH and DTFE methods to estimate the density fields (magenta and orange respectively). Results from the high-resolution skewer with $1024^3$ cells/particles are depicted in Fig.~\ref{fig:skewers}, left panel.}
\label{fig:skewer_lowres}
\end{figure}

At high resolution it is not easy to visually discern the differences between code outputs. To make these differences more obvious, Fig.~\ref{fig:skewer_lowres} shows the same low-density skewer as in Fig.~\ref{fig:skewers} to highlight the differences, i.e., the smoothing of small-scale fluctuations and the over-estimation of the density field in low-density regions using the SPH algorithm and the improvement due to the DTFE algorithm. Note that the DTFE density estimation-based results are much closer to the Nyx outputs at all resolutions (the SPH over-smoothing is very clear in the top two panels of the figure), but there remain differences in the details.

\subsection{Impact of the number of neighbors on the 1D \lya\ power spectra for \CRKHACC.sph}
\label{sec:nb}

\begin{figure*}
\centering
\includegraphics[width=0.49\textwidth]{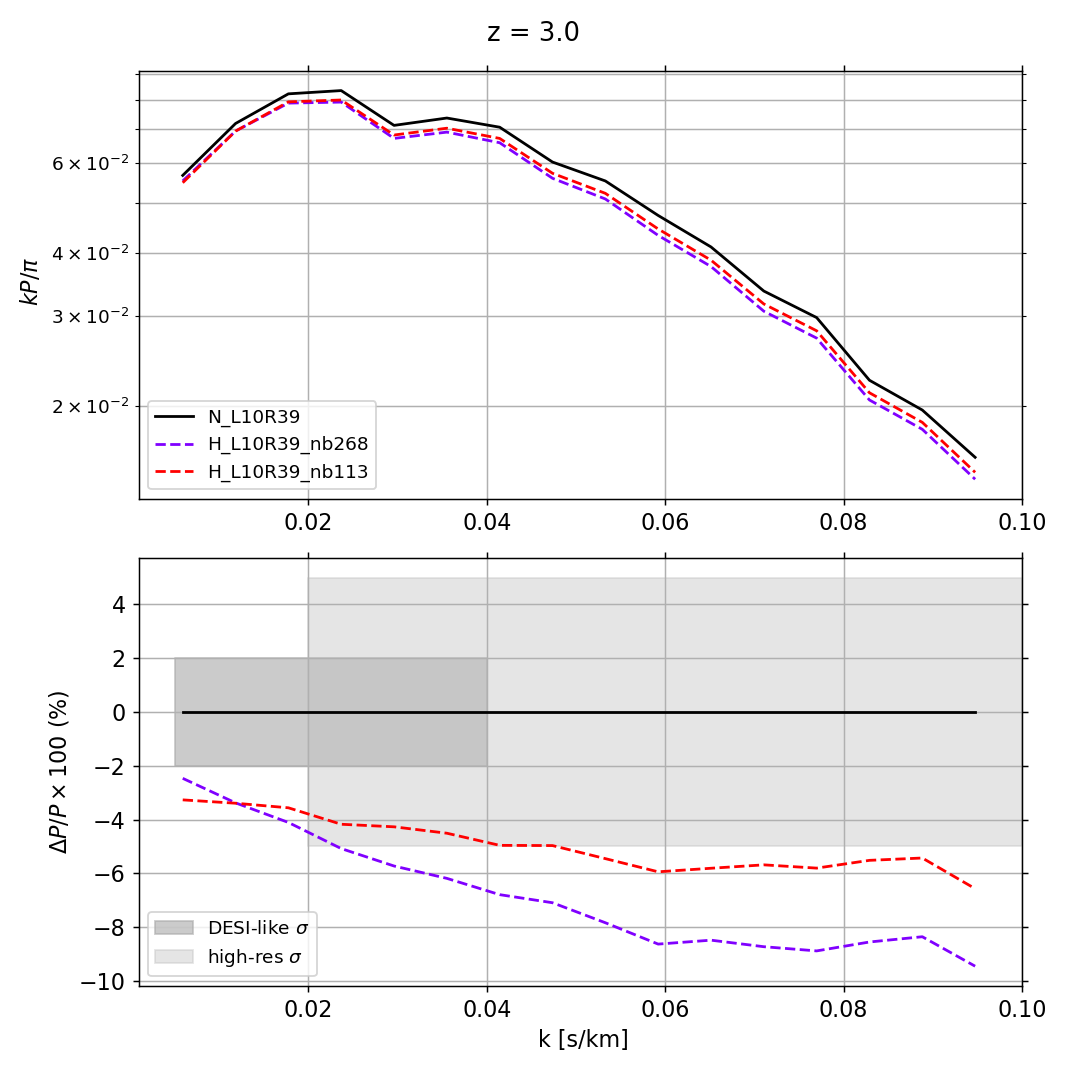}
\includegraphics[width=0.49\textwidth]{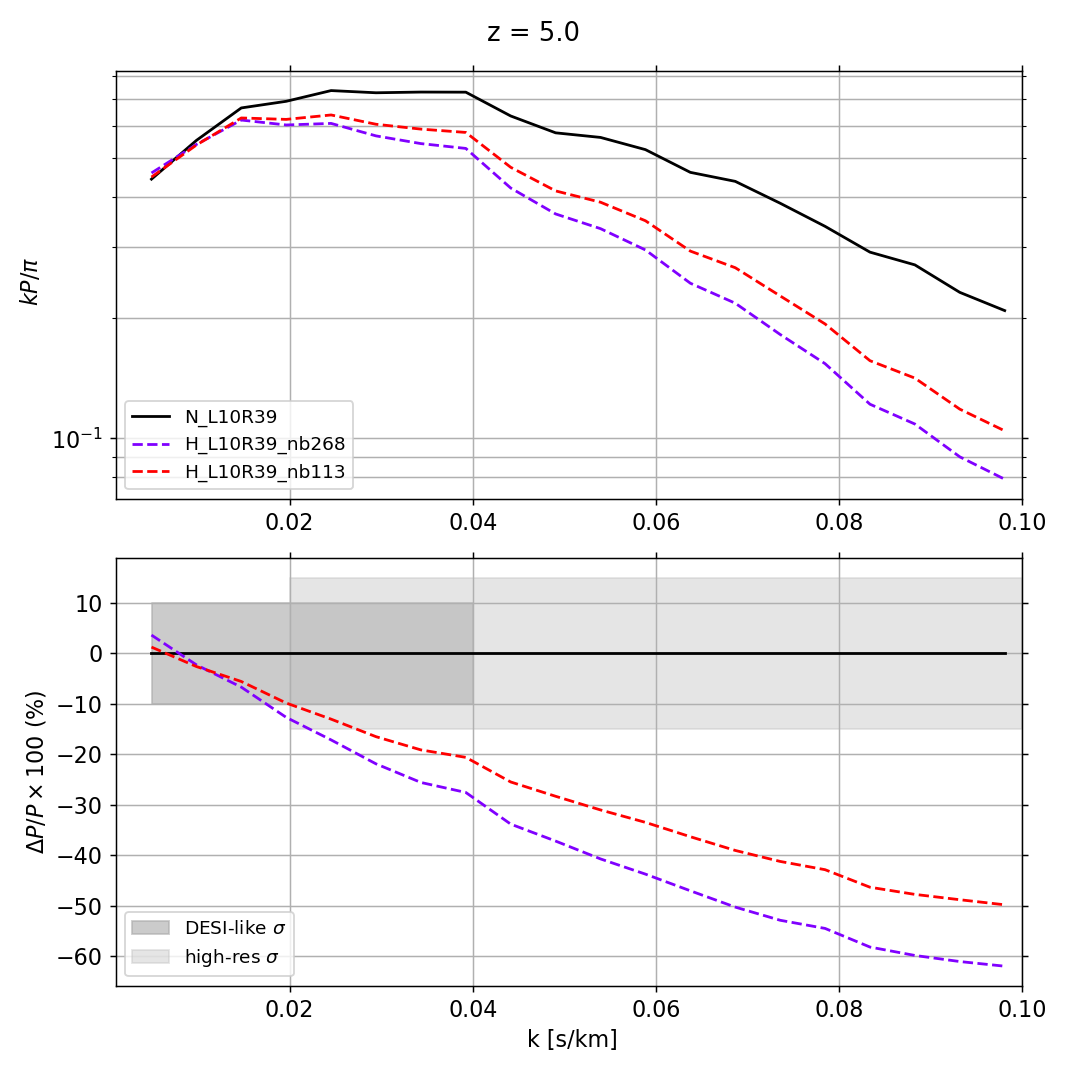}
\caption{1D Ly$\alpha$ power spectrum for Nyx (solid) and \CRKHACC.sph (dashed) at $z=3$ (left) and $z=5$ (right) in the 10~$\rm h^{-1}Mpc$ boxes. For \CRKHACC.sph, results are shown for different number of neighbors, $n_{\rm nb}$, contained in each smoothing sphere, with $n_{\rm nb} = 268$ and $n_{\rm nb} = 113$ in purple and red respectively. Middle and bottom panels show relative differences with respect to N\_L10R10, the Nyx most-resolved simulation with 10~$\rm h^{-1}kpc$ resolution. The shaded regions show $k$-averaged observational uncertainties of  DESI-like (darker gray) and high-resolution surveys respectively. These results can be contrasted with those in Fig.~\ref{fig:p1d}, which show the much-improved agreement when using the DTFE algorithm with \CRKHACC.}
\label{fig:nb}
\end{figure*}

We study the impact of SPH density estimation on the \lya\ 1D power spectrum by varying the the number of neighbors, $n_{\rm{nb}}$, for each SPH smoothing sphere. The results are shown in Fig.~\ref{fig:nb}. (See Sec.~\ref{sec:hacc} for a complete description of the parameter, $n_{\rm{nb}}.$). Lowering the number of neighbors, which increases the spatial resolution, makes the \CRKHACC power spectra prediction come in to slightly better agreement with Nyx, 2\% at $z=3$ and 10\% at  $z=5$. Reducing $n_{\rm{nb}}$ further does not improve the results as the reduction in smoothing is accompanied by an increase in noise leading to computational convergence issues in the high-order CRKSPH solver. The results in Fig.~\ref{fig:nb} can be contrasted with those in Fig.~\ref{fig:p1d}, which show the much-improved agreement when using the DTFE algorithm with \CRKHACC.

\subsection{Impact of low-density regions on the 1D \lya\ power spectra}
\label{sec:densitythresh}

\begin{figure*}
\centering
\includegraphics[width=0.49\textwidth]{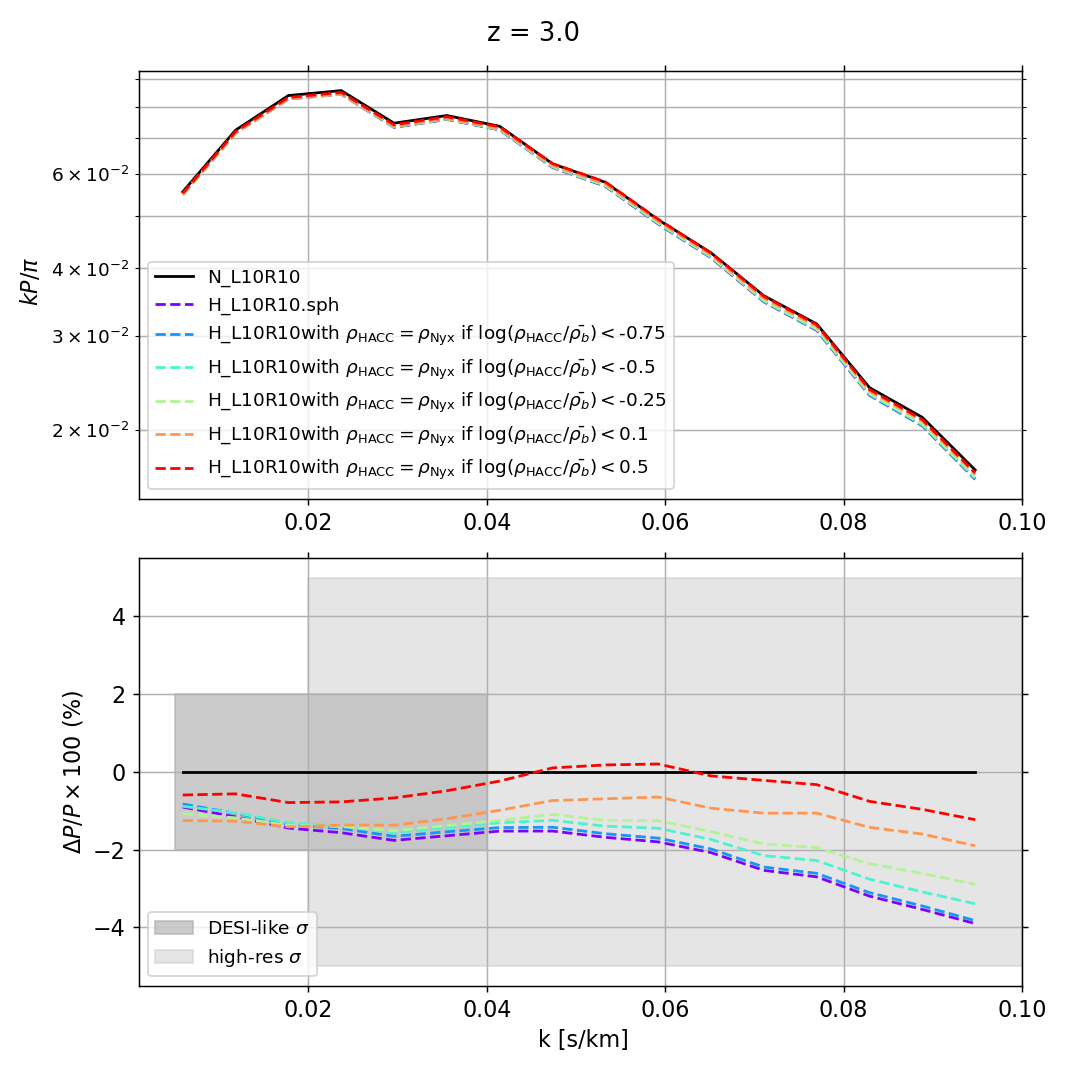}
\includegraphics[width=0.49\textwidth]{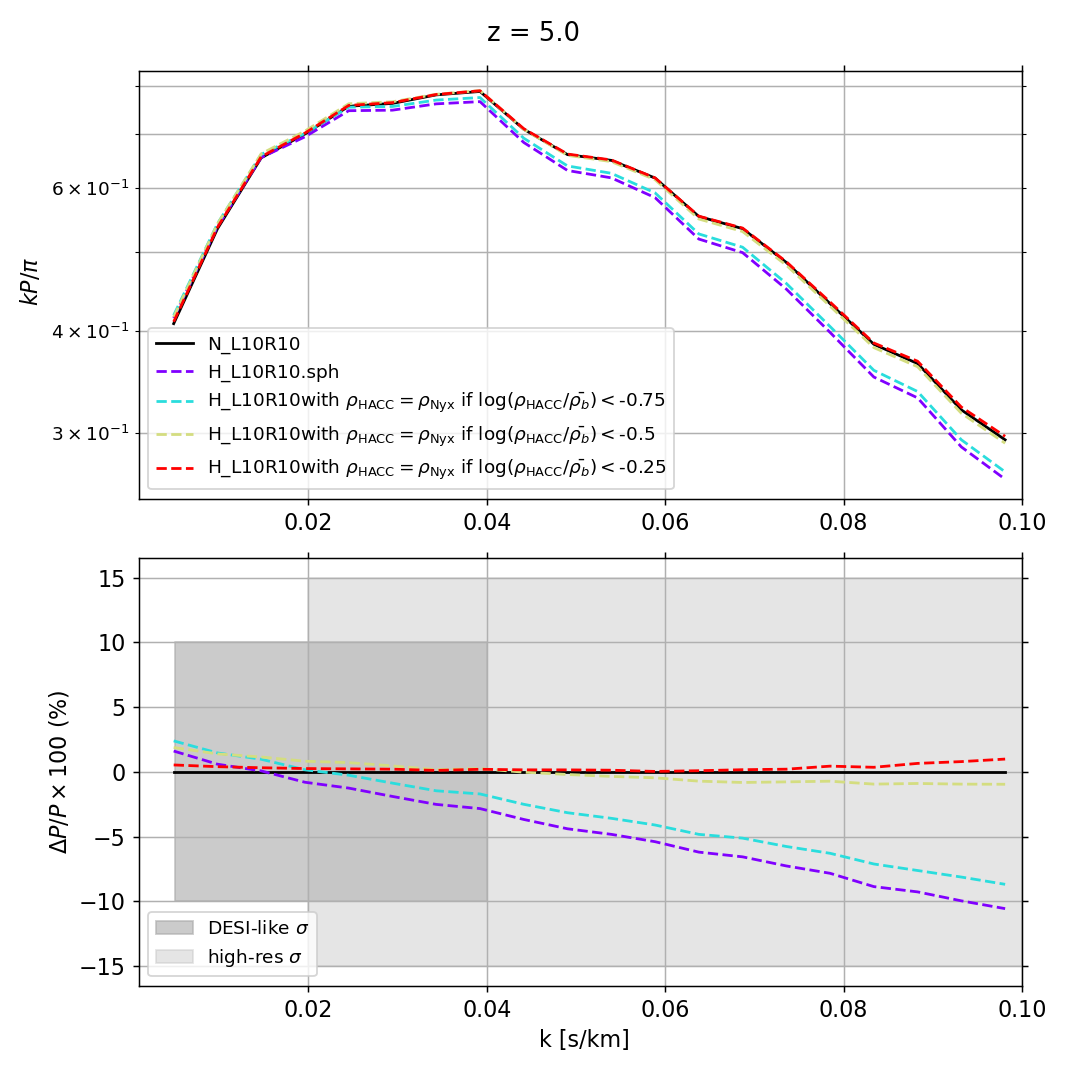}
\caption{1D Ly$\alpha$ power spectrum for Nyx (solid) and \CRKHACC.sph (dashed) at $z=3$ (left) and $z=5$ (right) in the 10 $\rm h^{-1}Mpc$ boxes. For \CRKHACC.sph, we use the Nyx density, $\rho_{\rm Nyx}$ instead of the \CRKHACC density, $\rho_{\rm HACC}$, when the latter is above $\rho_{\rm thresh}$ (in mean density units). Middle and bottom panels show relative differences with respect to N\_L10R10, the most-resolved Nyx simulation with 10~$\rm h^{-1}kpc$ resolution. The shaded regions show $k$-averaged observational uncertainties of  DESI-like (darker gray) and high-resolution (lighter gray) surveys respectively.}
\label{fig:densitythresh}
\end{figure*}

An investigation of the low-density behavior of the \lya\ 1D power spectrum from \CRKHACC.sph is shown in Fig.~\ref{fig:densitythresh}. As described in the main text (Section~\ref{comp_sph}), we carried out a number of tests and confirmed that the difference between the Nyx and \CRKHACC.sph arose from the low density regions. To see what regions were responsible for the discrepancy, we simply replaced the \CRKHACC.sph density field with the Nyx density field on the evaluation grid, when the density fell below a certain threshold value, taken to be a tunable parameter. For this test, we use a mesh with $N_{\rm grid} =   N_{\rm part}$ for the \CRKHACC outputs to match the Nyx mesh. We find that \CRKHACC.sph comes into increasing agreement as the parameter $\rho_{\rm thresh}$ is gradually increased; thus
highlighting that the main differences between the two codes originates from low-density regions. We find that replacing \CRKHACC.sph densities with Nyx densities for $\log(\rho_{\rm thresh}/\bar{\rho}_{b}) < -0.5~(0.5)$ is sufficient to reach sub-percent agreement between the two codes at $z=5~(z=3)$ over all scales. This sets the density scale to investigate the source of the possible discrepancy.

\begin{figure*}
\centering
\includegraphics[width=1.00\textwidth]{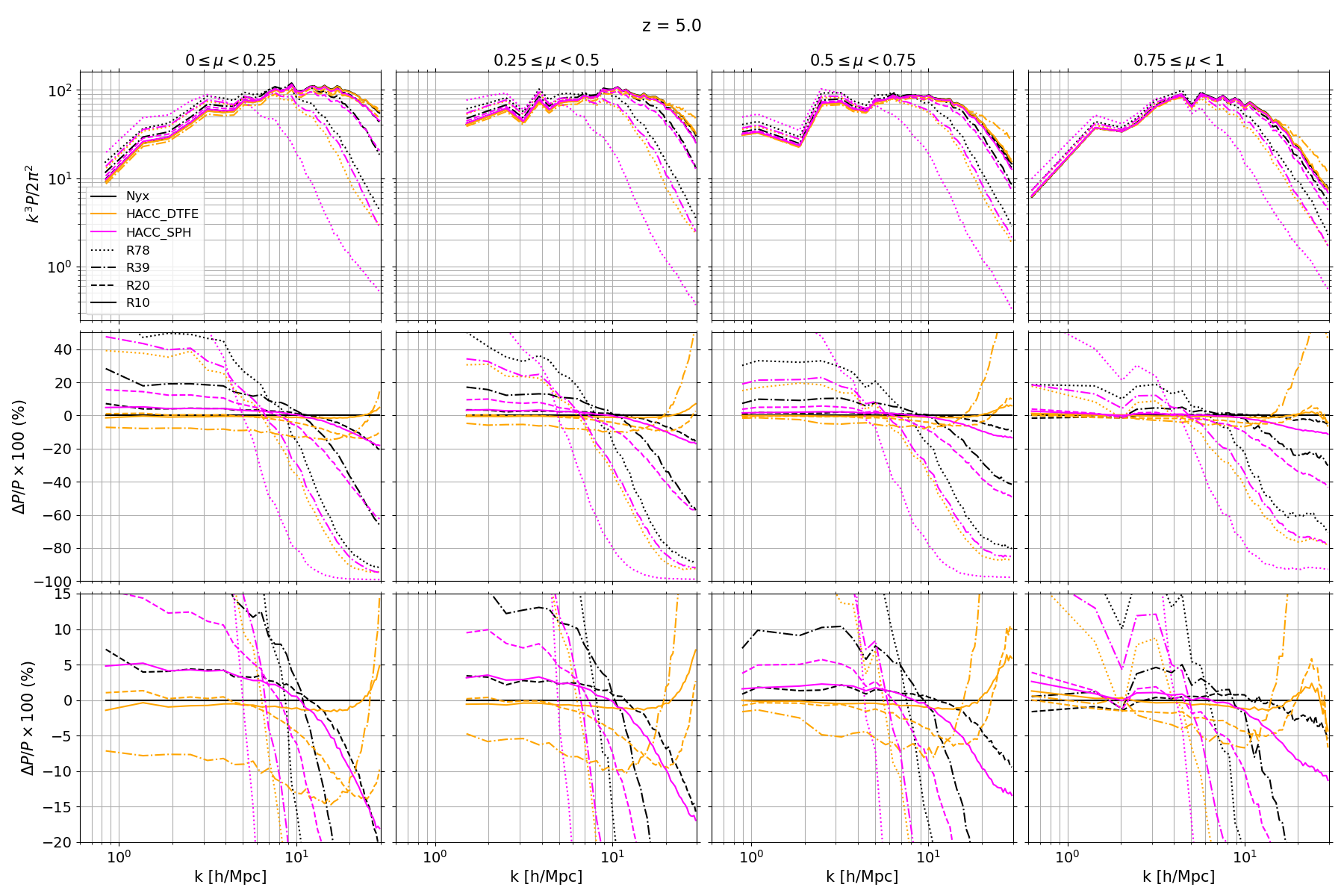}
\caption{3D Ly$\alpha$ power spectra at $z=5$ for Nyx (black), \CRKHACC.sph (magenta) and \CRKHACC.dtfe (orange) in 10~h$^{-1}$Mpc boxes at different spatial resolutions. Middle and bottom panels show relative differences with respect to N\_L10R10. The bottom panel shows a more restricted range on the $y$-axis to make it easier to distinguish the different results.}
\label{fig:p3d_highz}
\end{figure*}

\subsection{3D power spectrum at high redshifts}
\label{sec:p3dhighz}

Fig.~\ref{fig:p3d_highz} shows the 3D \lya\ power spectra at $z=5$ which can be contrasted with the $z=3$ results shown in Fig.~\ref{fig:p3d} (Sec.~\ref{sec:3dpspec}). The tendency of reduced error as one moves to the right (towards the LOS result) is less marked in this case compared to Fig.~\ref{fig:p3d}, due to the reduced thermal broadening at high redshifts. As at $z=3$, the power spectra from CRK-HACC.dtfe display a different sense of convergence as a function of $k$ compared to the Nyx and \CRKHACC.sph results, with less power at large scales and
higher power at small scales compared to the reference.

\subsection{Large box results}
\label{sec:largebox}

\begin{figure*}
\centering
\includegraphics[width=0.49\textwidth]{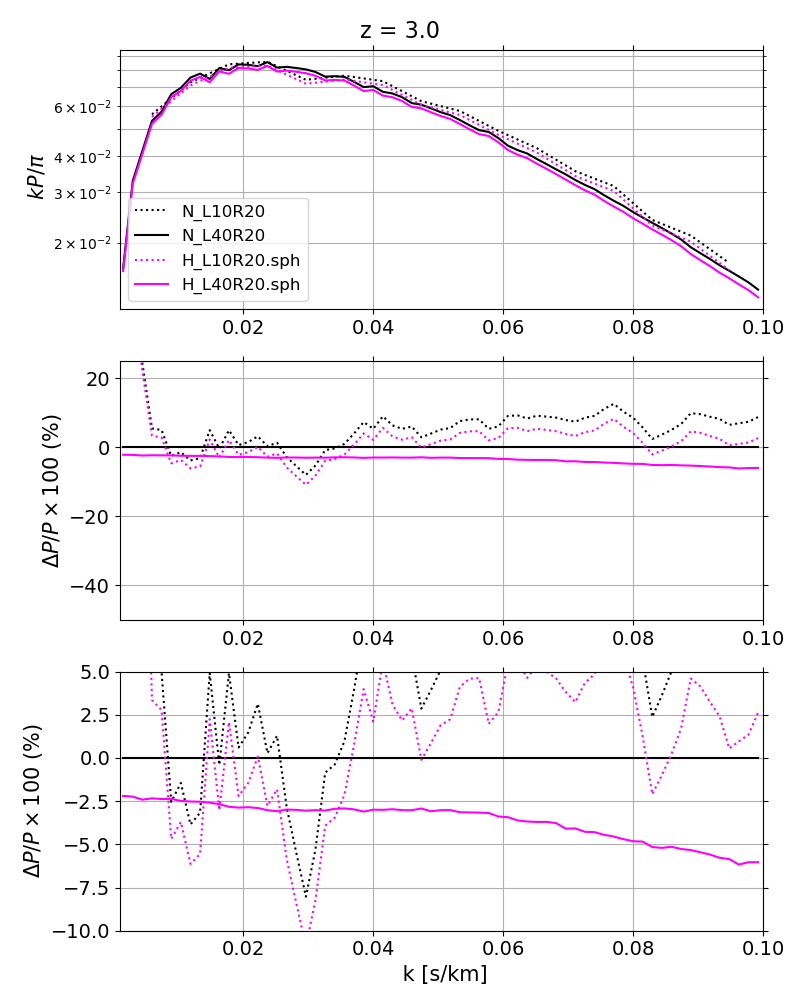}
\includegraphics[width=0.49\textwidth]{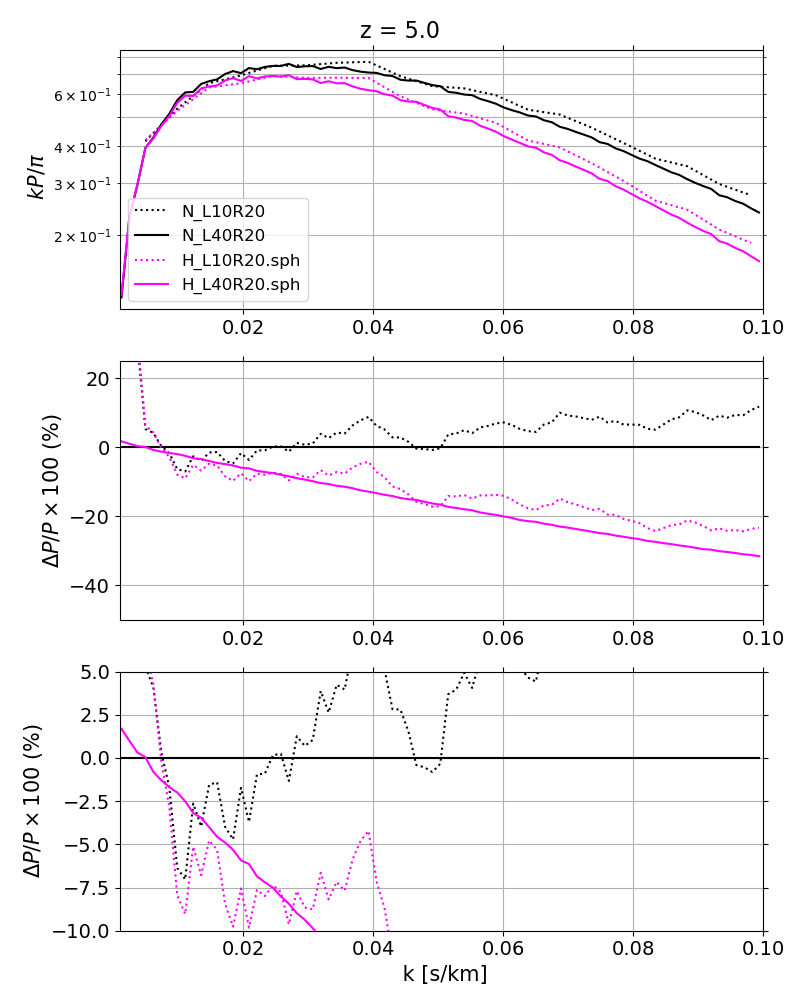}
\caption{1D Ly$\alpha$ power spectra for Nyx (black) and \CRKHACC.sph (magenta) at $z=3$ (left) and $z=5$ (right) in the 40~h$^{-1}$Mpc (solid) and 10~h$^{-1}$Mpc (dotted) boxes with an average resolution of 20 ~h$^{-1}$kpc. Middle and bottom panels show relative differences with respect to N\_L40R20. The bottom panel shows a more restricted range on the $y$-axis.}
\label{fig:p1d_boxsize}
\end{figure*}

\begin{figure*}
\centering
\includegraphics[width=0.95\textwidth]{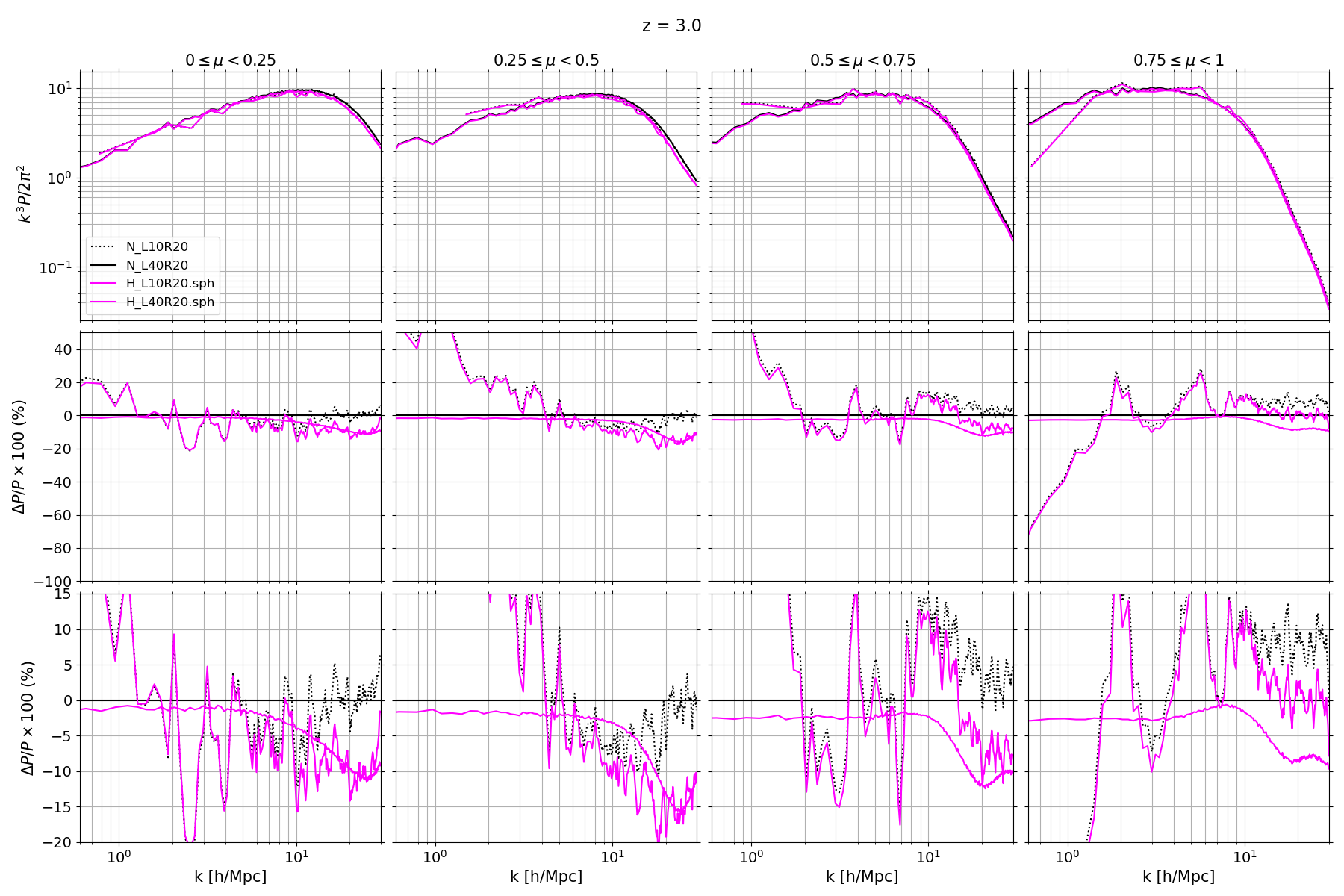}
\caption{
3D~Ly$\alpha$ power spectrum for Nyx (black) and \CRKHACC.sph (magenta) at $z=3$  in the 40~h$^{-1}$Mpc (solid) and 10~h$^{-1}$Mpc (dotted) boxes with an averaged resolution of 20~h$^{-1}$kpc. Middle and bottom panels show relative differences with respect to N\_L40R20. The bottom panel shows a more restricted range on the $y$-axis.
}
\label{fig:p3d_boxsize}
\end{figure*}

\begin{figure*}
\centering
\includegraphics[width=0.49\textwidth]{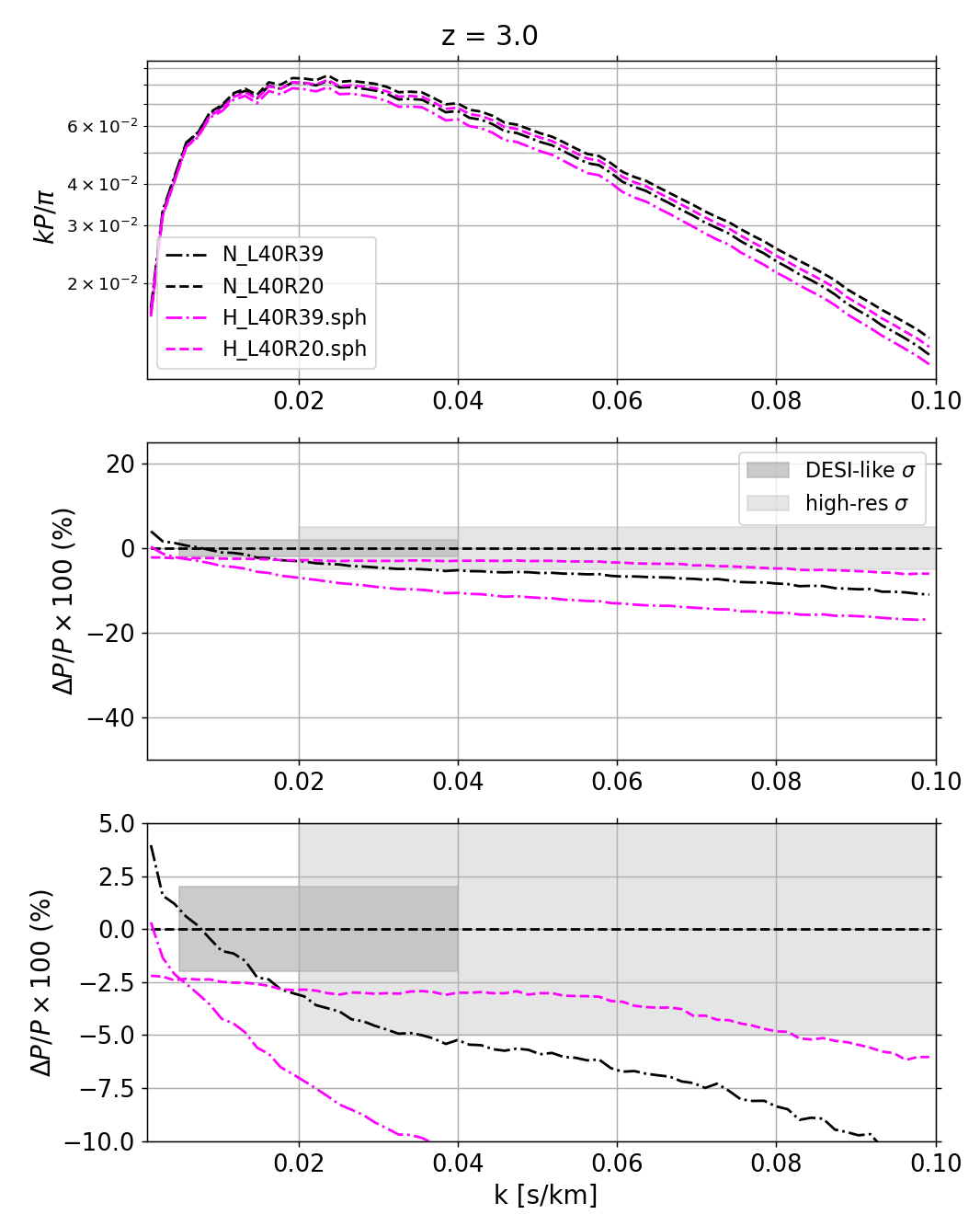}
\includegraphics[width=0.49\textwidth]{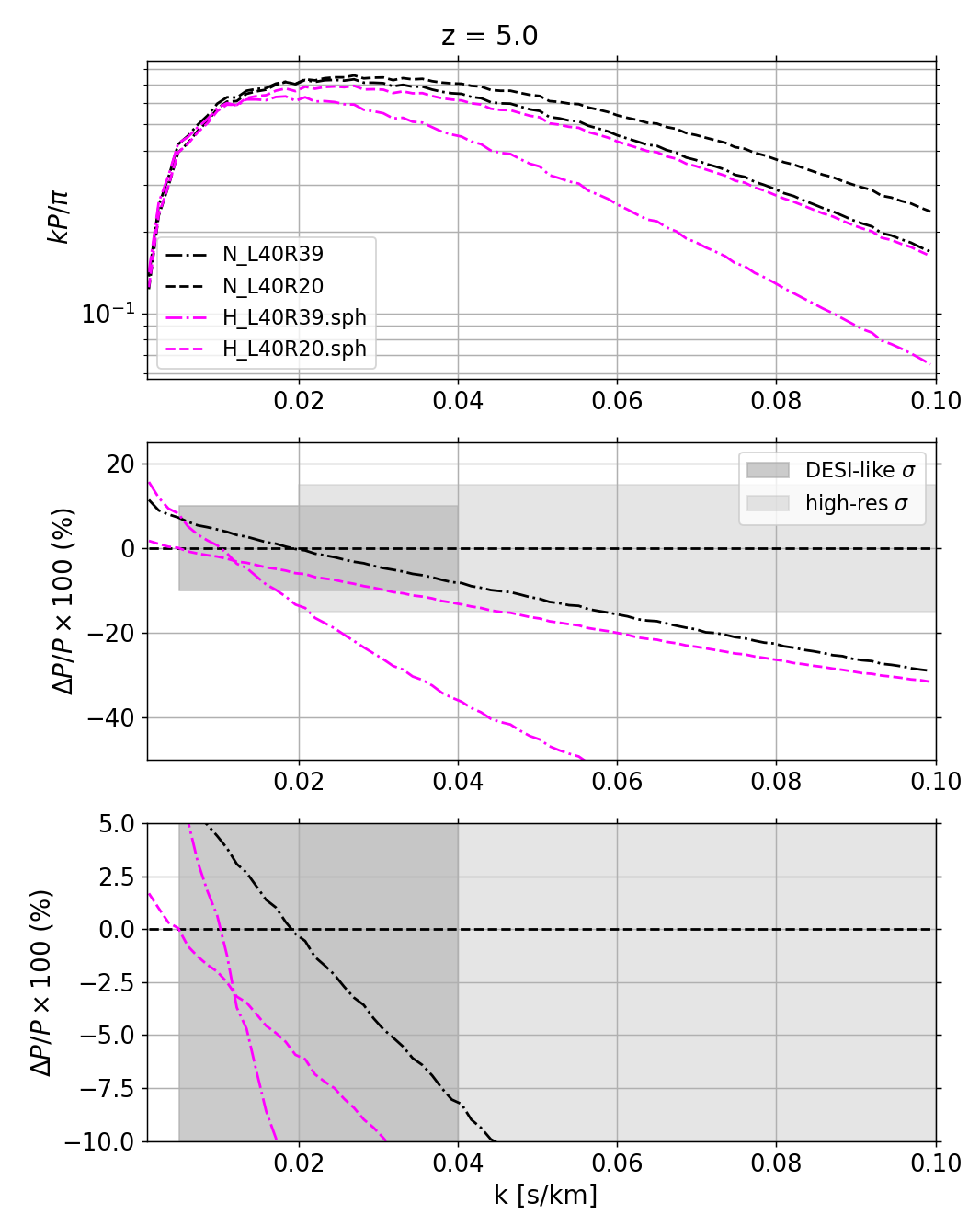}
\caption{1D Ly$\alpha$ power spectra for Nyx and \CRKHACC.sph at $z=3$ (left) and $z=5$ (right) in the 40~h$^{-1}$Mpc boxes at different spatial resolutions. Middle and bottom panels show relative differences with respect to N\_L40R10. Shaded regions show $k$-averaged observational uncertainties of DESI-like (darker gray) and high-resolution surveys (silver).}
\label{fig:p1d_L40}
\end{figure*}

Results from our investigation in box size effects are discussed here, adding to the presentation already provided in Sec.~\ref{sec:boxsize}. Fig.~\ref{fig:p1d_boxsize} and Fig.~\ref{fig:p3d_boxsize} show the impact of box size on the 1D  and 3D \lya\ power spectra respectively at $z=3$. The main purpose of showing these results is to demonstrate the consistent offset between the power spectra for both the Nyx and \CRKHACC.sph runs, which is due to the missing power in the smaller box, as mentioned already in Sec.~\ref{sec:boxsize}. Consequently, there is no additional relative correction needed for box size compensation. To complete the discussion, we also include data on the affect of spatial resolution in the larger 40~h$^{-1}$Mpc box: Fig.~\ref{fig:p1d_L40} shows the impact of varying resolution on the 1D \lya\ power spectra at $z=3$. The smooth behavior is consistent with the similar investigation carried out in Sec.~\ref{sec:richardson} for the 10~h$^{-1}$Mpc boxes.

\bsp	
\label{lastpage}
\end{document}